\documentclass{gGAF2e}

\usepackage{subfigure}
\usepackage{placeins}
\usepackage{float}
\usepackage{multiequations}
\newcommand*{\be}{\begin{equation}}
\newcommand*{\ee}{\end{equation}}
\newcommand*{\bse}{\begin{subequations}}
\newcommand*{\ese}{\end{subequations}}
\newcommand*{\bme}{\begin{multiequations}}
\newcommand*{\eme}{\end{multiequations}}
\newcommand*{\se}{\singleequation}

\newcommand*{\te}{\tripleequation}

\begin{document}

\jvol{00} \jnum{00} \jyear{2017} \jmonth{August}

\markboth{F. Wilson and T. Neukirch}{3D solutions of the magnetohydrostatic equations}

\title{Three-dimensional solutions of the magnetohydrostatic equations for rigidly rotating magnetospheres in cylindrical coordinates}

\author{F. Wilson$^{\ast}$\thanks{$^\ast$ Email: fw237@st-andrews.ac.uk
\vspace{6pt}} and T. Neukirch$^{\dag}$\thanks{$^\dag$ Email: tn3@st-andrews.ac.uk
\vspace{6pt}}\\\vspace{6pt}  School of Mathematics and Statistics, University of St Andrews, St Andrews, UK, KY16 9SS.\\\vspace{6pt}\received{\today}}

\maketitle

\begin{abstract}
We present new analytical three-dimensional solutions of the magnetohydrostatic equations, which are applicable to the co-rotating frame of reference outside a rigidly rotating cylindrical body, and have potential applications to planetary magnetospheres and stellar coronae. We consider the case with centrifugal force only, and use a transformation method in which the governing equation for the ``pseudo-potential" (from which the magnetic field can be calculated) becomes the Laplace partial differential equation. The new solutions extend the set of previously found solutions to those of a ``fractional multipole" nature, and offer wider possibilities for modelling than before. We consider some special cases, and present example solutions. 
\end{abstract}


\section{Introduction}

Analytical solutions of the magnetohydrostatic (MHS) equations without the assumption of spatial symmetries are very difficult to find; therefore, only a limited number of examples are known. One method for finding three-dimensional MHS equilibria, that is particularly appropriate for astrophysical applications, has been developed by Low and co-workers \citep{low-1982, low-1984,low-1985,Bogdan-1985,low-1991,low-1992,low-1993a,low-1993b,low-2005}, with additional contributions made by \citet{Neukirch-1995, Neukirch-1997}, \citet{Neukirch-1999}, \citet{Petrie-2000} and \citet{Rudenko-2001}.
Solutions found by this method have been applied to solar and astrophysical systems by many authors 
\citep[e.g.][]{Zhao-1993,Zhao-1994,Gibson-1995,Gibson-1996,Aulanier-1999,Zhao-2000,Ruan-2008,Lanza-2008, Lanza-2009,Wiegelmann-2015,Mactaggart-2016,Wiegelmann-2017}.

In order to allow for analytical solutions, the method relies on the presence of an external force; in the work referenced above, this external force is the gravitational force, either in Cartesian or spherical geometry. Another possibility for an external force is the centrifugal force that is present in the co-rotating frame of reference of a rigidly rotating plasma system \citep[e.g][]{low-1991}. The corresponding theory and a few example solutions have been described by \citet{Neukirch-2009b}. Further work by \citet{Al-Salti-2010a} and \citet{Al-Salti-2010b} has extended this to cases where the external force is a combination of centrifugal and gravitational forces. In these cases, however, numerical methods are needed to obtain solutions.

\citet{Neukirch-2009b} only touched upon a few possible special cases to find illustrative examples of solutions. The purpose of the work in this paper is to extend some of the ideas formulated by \citet{Neukirch-2009b} and to show that it is possible to find a far more general class of analytical solutions. We will use the same geometrical set-up as  \citet{Neukirch-2009b} by assuming a rigidly rotating cylindrical central body, and will focus on a particular class of equilibria found by \citet{Neukirch-2009b} (using a transformation method), discussing how this class of solutions can be extended considerably.

The structure of the paper is as follows: in section \ref{sec:theory}, we describe the basic theoretical approach, and previous work by \citet{Neukirch-2009b}. In section \ref{sec:cyl_new}, we discuss further cylindrical MHS solutions, and present some examples. In section \ref{sec:summary}, we give a summary and conclusions.

\section{Theoretical framework}

\label{sec:theory} 

\subsection{Brief overview of general theory}

In this section, we summarise the general theory developed by \citet{low-1985,low-1991,low-1992,low-1993a,low-1993b} for calculating a particular class of MHS equilibria. In this brief overview, we do not refer to a specific coordinate system, and hence the theory is applicable to various different cases 
\citep[e.g.][]{Neukirch-2009b,Al-Salti-2010a, Al-Salti-2010b}. However, later in this paper we will focus on rigidly rotating cylindrical bodies, as discussed by \citet{Neukirch-2009b}.

The MHS equations, in the frame of reference co-rotating with the rigidly rotating body, have the form \citep[e.g.][]{mestel2003}
\bme
\label{force-ampere-divB}
\be
\se
\bm{j}\times\bm{B}-\bm\nabla{p}-\rho\bm\nabla{V}=\bm{0}\,,
\ee
\vskip -6mm
\be
\bm\nabla\times\bm{B}=\mu_0\bm{j}\,,\qquad\qquad\qquad \bm \nabla{\bm\cdot}\bm{B}=0\,.
\ee
\eme
for current density $\bm{j}$, magnetic field $\bm{B}$, pressure $p$, density $\rho$, and potential $V$.  

In Low's method \citep[e.g.][]{low-1991}, it is assumed that the current density can be written in terms of the potential $V$ and a free function $F$ as
\begin{equation}
 \bm{j}=\frac{1}{\mu_0}\left(\bm\nabla{F}\times\bm\nabla{V}\right),\label{j_assumption}
\end{equation}
so that $\bm\nabla{\bm\cdot}\bm{j}=0$. This choice gives a current density that is perpendicular to the gradient of the potential and, although restrictive, is made to allow integrability of the force-balance equation.

Assuming that $F$ has the form
\begin{equation}
 F(V,\bm{B})=\kappa(V)\bm{B}{\bm\cdot}\bm\nabla{V}\label{F form},
\end{equation}
for a free function $\kappa(V)$ (that describes the deviation from a potential field), then gives the magnetic field (using Amp\'{e}re's law (\ref{force-ampere-divB}b))
as
\begin{equation}
 \bm{B}=\bm\nabla{U}+f_\kappa\left(\bm\nabla{U}{\bm\cdot}\bm\nabla{V}\right)\bm\nabla{V},\label{B}
\end{equation}
where
\begin{equation}
 f_\kappa=\frac{\kappa(V)}{1-\kappa(V)\left(\bm\nabla{V}\right)^2}.
\end{equation}
We refer to the unknown function $U$ in (\ref{B}) as a ``pseudo-potential", from which we can calculate the magnetic field; it can be found by substituting the expression for $\bm{B}$ from (\ref{B}) into the solenoidal constraint (\ref{force-ampere-divB}c),
which gives the governing equation
\begin{equation}
\bm\nabla{\bm\cdot}\left(\bm\nabla{U}+f_\kappa\left(\bm\nabla{U}{\bm\cdot}\bm\nabla{V}\right)\bm\nabla{V}\right)=0,\label{divb} 
\end{equation}
which can be written in the form
\begin{equation}
 \bm\nabla{\bm\cdot}\left(\bm{\sf M}\bm\nabla{U}\right)=0,\label{pde}
\end{equation}
for the $3\times3$ matrix $\bm{\sf M}$ given by
\begin{equation}
 \bm{\sf M}=\bm{\sf I}+f_\kappa\bm\nabla{V}\,\bm\nabla{V},
\end{equation}
where $\bm{\sf I}$ is the $3\times3$ unit matrix. Equation (\ref{divb}), or equivalently (\ref{pde}), is an elliptic partial differential equation provided $1-\kappa(V)\left(\bm\nabla{V}\right)^2>0$, otherwise it is hyperbolic \citep[e.g.][]{Al-Salti-2010a,Al-Salti-2010b}. The particular choice of $F$ given in (\ref{F form}) was made because it renders the fundamental equation (\ref{divb}), or (\ref{pde}), linear. For an example of a choice of $F$ that results in a nonlinear governing equation \citep[see][]{Neukirch-1997}.

Following the theory as described by, e.g., \citet{low-1991} or \citet{Neukirch-2009b}, the pressure and density are given by
\begin{subequations}
\begin{eqnarray}
 p&=&p_0(V)-\frac{F^2}{2\mu_0\kappa(V)},\label{p0}\\
\rho&=&-\left(\frac{\partial p}{\partial V}\right)_F+\frac{1}{\mu_0}(\bm{B}{\bm\cdot}\bm\nabla F)
\label{density2},
\end{eqnarray}
\end{subequations}
where $p_0(V)$ is a
hydrostatic background pressure,
which should be chosen appropriately to ensure that both the pressure and density are positive. 
The final expressions for the pressure and density can then be obtained by substituting back for $F$ into (\ref{p0},b)
(using (\ref{F form})), which gives
\begin{subequations}
\begin{eqnarray}
 p&=&p_0(V)-\frac{1}{2\mu_0}\kappa(V)(\bm{B}{\bm\cdot}\bm\nabla{V})^2\label{pressure},\\
\rho&=&-\frac{\mathrm{d}p_o}{\mathrm{d}V}+\frac{1}{2\mu_0}\frac{\mathrm{d}\kappa}{\mathrm{d}V}(\bm{B}{\bm\cdot}\bm\nabla{V})^2+\frac{1}{\mu_0}\kappa(V)\bm{B}{\bm\cdot}\bm\nabla(\bm{B}{\bm\cdot}\bm\nabla{V})
\label{density}.
\end{eqnarray}
\end{subequations}

The system of equations (\ref{force-ampere-divB}a-c)
can, finally, be closed by assuming a suitable equation of state. For example, for an ideal gas we have
\begin{equation}
 T=\frac{\mu_m p}{R_0\rho}
\end{equation}
for universal gas constant $R_0$ and mean molecular weight $\mu_m$.

\subsection{Rigidly rotating systems in cylindrical geometry}

As stated above, we aim to extend the solution class found by \citet{Neukirch-2009b} using cylindrical coordinates.
\citet{Neukirch-2009b} considered the somewhat artificial case of an infinitely long cylindrical body of radius $R_c$ rotating rigidly about its symmetry axis with a constant angular velocity $\varOmega$. Defining a cylindrical coordinate system $(\varpi,\phi,z)$ in the co-rotating frame of reference and ignoring gravity, the potential $V$ is the centrifugal potential, given by
\begin{equation}
V=-\,\tfrac{1}{2}\varOmega^2\varpi^2.
\label{centrifugal}
\end{equation}
We can simplify the mathematics of this problem by defining 
\begin{equation}
 \xi(V)=\kappa(V)\left(\bm\nabla{V}\right)^2,\label{xi_def}
\end{equation}
where $\left(\bm\nabla{V}\right)^2=-\varOmega^4\varpi^2$ for the centrifugal potential (\ref{centrifugal}). Since $V$ depends only on $\varpi$, $\xi(V)\equiv\xi(\varpi)$ in this case.
The governing equation for the ``pseudo-potential" $U$ (see (\ref{divb})) is then given in cylindrical coordinates as
\begin{equation}
\frac{1}{{\varpi}}\frac{\partial}{\partial {\varpi}}\left(\frac{\varpi}{1-\xi(\varpi)}\frac{\partial U}{\partial{\varpi}}\right)+\frac{1}{{\varpi}^2}\frac{\partial^2U}{\partial{\phi}^2}+\frac{\partial^2U}{\partial z^2}=0,\label{cyl_eq}
\end{equation}
and the magnetic field components can be calculated from (\ref{B}) as
\bme
\label{br-bphi-bz}
\te
\be
 B_\varpi=\frac{1}{1-\xi(\varpi)}\frac{\partial U}{\partial\varpi}\,,\qquad\qquad
B_\phi=\frac{1}{\varpi}\frac{\partial U}{\partial\phi}\,,\qquad\qquad
B_z=\frac{\partial U}{\partial z}\,.
\ee
\eme
The pressure and density expressions (\ref{pressure}) and (\ref{density}) can also be rewritten as
\begin{subequations}
\begin{eqnarray}
p&=&\bar{p}_0(\varpi)-\frac{\xi(\varpi)}{2\mu_0}B_\varpi^2\label{pressure_diff},\\
\rho&=&\frac{1}{\varOmega^2 \varpi}\Bigg[\frac{\mathrm{d}\bar{p}_o}{\mathrm{d}\varpi}-\frac{1}{2\mu_0}\frac{1}{(1-\xi(\varpi))^2}\frac{\mathrm{d}\xi}{\mathrm{d} \varpi}\left(\frac{\partial U}{\partial \varpi}\right)^2-\frac{1}{\mu_0}\xi(\varpi)(\bm{B}{\bm\cdot}\bm\nabla)B_\varpi\Bigg],
\end{eqnarray}
\end{subequations}
where we have corrected a typographical error in the work by \citet{Neukirch-2009b}. As previously discussed, the background pressure $\bar{p}_0(\varpi)$ must be chosen in such a way that both the pressure and density are positive. 

Using separation of variables, the general solution of (\ref{cyl_eq}) can be expressed as
\begin{eqnarray}
U(\varpi,\phi,z)&=&\sum_{m=-\infty}^{\infty}\exp({\mathrm i}m\phi)\int_{-\infty}^{\infty}\exp({\mathrm i}kz)\left[A_m(k)F_{mk}^{1}(\varpi)+B_m(k)F_{mk}^2(\varpi)\right]\mathrm{d}k,\qquad\qquad
\end{eqnarray}
for, generally, complex coefficients $A_m(k)$ and $B_m(k)$, which are determined by the boundary conditions. 

\section{The constant $\xi$ case}

\label{sec:cyl_new}

We will focus on the simplest case, $\xi=\xi_0$, for constant $0<\xi_0<1$. Using the separation of variables method of the previous section, we find that the functions $F_{mk}^i$ are modified Bessel functions \citep{Neukirch-2009b}. In this simplest case, however, it also is possible to find solutions of (\ref{cyl_eq}) by using the coordinate transformation
\bme
\label{transformation}
\be
\bar{\varpi}=\sqrt{1-\xi_0}\varpi, \hskip 20mm  \bar{\phi}=\frac{\phi}{\sqrt{1-\xi_0}},
\ee
\eme
as pointed out by \citet{Neukirch-2009b}. We will focus on this method for the remainder of the paper. The transformation gives (\ref{cyl_eq}) as the Laplace equation in the  transformed cylindrical coordinates $(\bar{\varpi},\bar{\phi},z)$;
\begin{equation}
\frac{1}{\bar{\varpi}}\frac{\partial}{\partial \bar{\varpi}}\left(\bar{\varpi}\frac{\partial U}{\partial\bar{\varpi}}\right)+\frac{1}{\bar{\varpi}^2}\frac{\partial^2U}{\partial\bar{\phi}^2}+\frac{\partial^2U}{\partial z^2}=0,\label{cyl_laplace}
\end{equation}
which, in principle, allows solutions of the Laplace equation to be transformed into solutions of (\ref{cyl_eq}) for $\xi(\varpi)=\xi_0$.

One important aspect to note is that, for a valid solution, $U$ must be $2\pi$-periodic in the untransformed azimuthal coordinate $\phi$. 
To achieve this, \citet{Neukirch-2009b} chose $\xi_0 = 1 - n^{-2}$, with $n$ an integer $\ge 2$. This implies that any $\bar{\phi}$ dependence of a solution of (\ref{cyl_laplace}) will be transformed into an $n\phi$ dependence and, hence, the lowest order dependence on the untransformed azimuthal coordinate, $\phi$, is a $2\phi$ dependence. Additionally, superposition of solutions for different $n$ values is not possible, because they would correspond to solutions for different values of $\xi_0$, and it is also unsatisfactory that the method cannot be used for arbitrary values of $\xi_0 < 1$.

We will now demonstrate how to generalise the method using the transformation (\ref{transformation}) in such a way that the restrictions we just mentioned no longer apply, which allows us to find more general MHS equilibrium solutions for the rigidly rotating cylinder case with $\xi(\varpi)=\xi_0$. We will make use of the fact that we are free to solve Laplace's equation (\ref{cyl_laplace}) in any coordinate system of our choice. We define
the ``pseudo-spherical" coordinates $(\bar{r},\bar{\theta},\bar{\phi})$ in terms of our transformed cylindrical coordinates $(\bar{\varpi},\bar{\phi},z)$, where
\bme
\label{pseudo1-2}
\be
\bar{r}^2=\bar{\varpi}^2+z^2,   \hskip 20mm
\cos\bar{\theta}=\frac{z}{\bar{r}}.
\ee
\eme
The Laplace equation (\ref{cyl_laplace}) is then given by
\begin{eqnarray}
\frac{1}{\bar{r}^2}\frac{\partial}{\partial\bar{r}}\left(\bar{r}^2\frac{\partial U}{\partial\bar{r}}\right)+\frac{1}{\bar{r}^2\sin\bar{\theta}}\frac{\partial}{\partial\bar{\theta}}\left(\sin\bar{\theta}\frac{\partial U}{\partial\bar{\theta}}\right)
+\frac{1}{\bar{r}^2\sin\bar{\theta}}\frac{\partial^2 U}{\partial\bar{\phi}^2}=0.\label{laplace_pseudo}
\end{eqnarray}
This equation can be solved by using the separation of variables
\begin{equation}
 U\left(\bar{r},\bar{\theta},\bar{\phi}\right)=R\left(\bar{r}\right)\varTheta\left(\bar{\theta}\right)\varPhi\left(\bar{\phi}\right),
\end{equation}
which gives solutions of the form
\begin{eqnarray}
 U(\bar{r},\bar{\theta},\bar{\phi})&=&\bigl(A_{\mu\nu}\bar{r}^{-(\nu+1)}+B_{\mu\nu}\bar{r}^{\nu}\bigr)\bigl[C_{\mu\nu}{\mathrm P}_\nu^\mu(\cos\bar{\theta})+D_{\mu\nu}{\mathrm P}_\nu^\mu(-\cos\bar{\theta})\bigr]\exp({\mathrm i}\mu\bar{\phi}),\qquad\qquad\label{laplace_sol}
\end{eqnarray}
where the $\mathrm{P}^\mu_\nu$ functions are associated Legendre functions of the first kind \citep{Abramowitz-1964, NIST:DLMF}, and  $A_{\mu\nu}$, $B_{\mu\nu}$, $C_{\mu\nu}$ and $D_{\mu\nu}$ are constants that are determined by the boundary conditions. For example, if we impose the condition that $U$ tends to zero as $\bar{r} \to \infty$, then the constant $B_{\mu\nu}=0$ (if we assume $\nu>0$). In general, both $\mu$ and $\nu$ can be non-integers, with $\mu$ the order and $\nu$ the degree of the Legendre functions.
If $\mu$ and $\nu$ are not integers, the solutions are reminiscent of fractional multipoles (see, e.g., \cite{Engheta-1996,Debnath-2003}). Note that, in the solutions (\ref{laplace_sol}), we have chosen the two associated Legendre functions ${\mathrm P}_\nu^\mu(\cos\bar{\theta})$ and ${\mathrm P}_\nu^\mu(-\cos\bar{\theta})$; these are linearly independent unless both $\mu$ and $\nu$ are integers.

We can constrain the possible values for $\mu$ by the requirement that the solution must be $2\pi$-periodic in the untransformed azimuthal coordinate $\phi$. If we assume that we have chosen a value for $\xi_0$ (with $0 < \xi_0 < 1$), this gives 
\begin{equation}
 \mu=m \sqrt{1-\xi_0}\label{muuuu},
\end{equation}
for integer $m$, with no further restriction imposed on the value for $\xi_0$. In particular, unlike for the previously discussed solutions \citep{Neukirch-2009b}, we are no 
longer restricted to a lowest order dependence of $2\phi$; we are free to choose $m=1$.

Since we are concerned with solutions outside a cylindrical body, we will write the solution (\ref{laplace_sol}) (taking $B_{\mu\nu}=0$) in terms of our transformed cylindrical coordinates as
\begin{eqnarray}
 U(\bar{\varpi},\bar{\phi},z)&=&\frac{A_{\mu\nu}\exp({\mathrm i}\mu\bar{\phi})}{\left(\bar{\varpi}^2+z^2\right)^{(\nu+1)/2}} \left[C_{\mu\nu}{\mathrm P}_\nu^\mu\!\left(\frac{z}{\sqrt{\bar{\varpi}^2+z^2}}\right)+D_{\mu\nu}{\mathrm P}_\nu^\mu\!\left(-\frac{z}{\sqrt{\bar{\varpi}^2+z^2}}\right)\right].\qquad\qquad\label{laplace_sol_cyl}
\end{eqnarray}

We note that the Legendre functions can be singular when their argument equals $\pm1$; from (\ref{laplace_sol_cyl}), this corresponds to $\bar{\varpi}=0$ and $z\to\pm\infty$. Since we are interested only in the solution outside the central cylindrical body, points with $\bar{\varpi}=0$, i.e. on the rotation axis, do not present a problem, since they are hidden inside the cylinder (note, however, that this would be problematic in the case of a central rigidly rotating spherical body). By looking at the asymptotic behaviour of the Legendre functions, it is possible to show that, for a given choice of $\nu$, there exists a range of permitted $\mu$ values for which the ``pseudo-potential" $U(\bar{\varpi},\bar{\phi},z)$, and all of the terms in the force balance equation (\ref{force-ampere-divB}a),
vanish as $z\to\pm\infty$  (i.e. on the upper and lower boundaries of the cylindrical domain), as required for a physically reasonable solution. For a further discussion of this point, see the appendix.

The governing equation for the ``pseudo-potential" $U$ (see (\ref{laplace_pseudo})) of course ensures that the magnetic field is divergence free. For a valid solution, however, we have the additional requirement that the integral of $\bm\nabla{\bm\cdot}\bm{B}$ over the cylindrical volume, $V_c$, must vanish or, equivalently, that the flux through the cylindrical surface, $S_c$, vanishes;
\begin{equation}
 \int_{V_c}\left(\bm\nabla{\bm\cdot}\bm{B}\right)\mathrm{d}V=\int_{S_c}\bm{B}{\bm\cdot}\mathrm{\bm{d}}\bm{S}=0.\label{gauss_thm}
\end{equation}
We will discuss this further for the particular form of solution that we consider in section \ref{sec:special}.

\subsection{Special case: $\mu=-\nu$} 

\label{sec:special}

As discussed in section \ref{sec:cyl_new}, the Legendre functions contained in the solution (\ref{laplace_sol_cyl}) can generally have singularities, which places restrictions on the parameters $\mu$ and $\nu$. The following closed form exists for the Legendre functions, however, for the case $\mu=-\nu$ \citep{Abramowitz-1964, NIST:DLMF}; 
\begin{eqnarray}
 {\mathrm P}_{\nu}^{-\nu}(x)=\frac{2^{-\nu}\left(1-x^2\right)^{\nu/2}}{\Gamma(\nu+1)}.\label{mu=-nu}
\end{eqnarray}
Note that, under these conditions, $\mathrm{P}$ is an even function, and so the two Legendre functions in (\ref{laplace_sol_cyl}) are equal. Using (\ref{mu=-nu}), we can, for a given $\nu$, consider a single solution of (\ref{laplace_sol_cyl}), in terms of the original (untransformed) cylindrical coordinates $\left(\varpi,\phi,z\right)$, as
\begin{eqnarray}
  U({\varpi},{\phi},z)&=&U_0\frac{2^{-\nu}\left(1-\xi_0\right)^{\nu/2}}{\Gamma(\nu+1)}\frac{{\varpi}^{\nu}\cos(\nu{\phi}/\sqrt{1-\xi_0})}{\left[\left(1-\xi_0\right){\varpi}^2+z^2\right]^{\nu+1/2}},\label{usol}
\end{eqnarray}
for constant $U_0$. For illustrative purposes later, when we superpose the new solutions with a magnetic dipole solution, we normalise $\varpi$ and $z$ by the cylinder radius $R_c$ and choose 
\begin{equation}
 U_0=\frac{\mu_0m_d}{4\pi}\frac{A_\nu R_{c}^{\nu-1}\Gamma(\nu+1)}{2^{-\nu}(1-\xi_0)^{\nu/2}},\label{u0}
\end{equation}
where $m_d$ is a parameter with dimensions of a dipole moment. We then normalise $U$ by $\mu_0m_d/(4\pi R_c^{2})$, which gives the normalised ``pseudo potential" as
\begin{eqnarray}
  U({\varpi},{\phi},z)&=&A_\nu\frac{{\varpi}^{\nu}\cos(\nu{\phi}/\sqrt{1-\xi_0})}{\left[\left(1-\xi_0\right){\varpi}^2+z^2\right]^{\nu+1/2}},\label{usol}
\end{eqnarray}
where $\varpi$ and $z$ now refer to normalised quantities. It can be seen that, provided $\nu>-1/2$, the solution will tend to zero in the limit $z\to\pm\infty$ (and $\varpi\to\infty$, which must also be true for a physically viable solution).

Using (\ref{br-bphi-bz}a-c),
the magnetic field components are given by
\begin{subequations}
\begin{eqnarray}
  B_\varpi&=&A_\nu\frac{\cos\left(\nu\phi/\sqrt{1-\xi_0}\right)}{1-\xi_0}\left[\frac{\nu\varpi^{\nu-1}}{\left[(1-\xi_0)\varpi^2+z^2\right]^{\nu+1/2}}-\frac{(1-\xi_0)(2\nu+1)\varpi^{\nu+1}}{\left[(1-\xi_0)\varpi^2+z^2\right]^{\nu+3/2}}\right]\label{br2}\label{B_varpi},\\
 B_\phi&=&-\frac{A_\nu\nu}{\sqrt{1-\xi_0}}\frac{\varpi^{\nu-1}\sin\left(\nu\phi/\sqrt{1-\xi_0}\right)}{\left[(1-\xi_0)\varpi^2+z^2\right]^{\nu+1/2}}\label{bphi2}\label{B_phi},\\
 B_z&=&-A_\nu\left(2\nu+1\right)\frac{z\varpi^{\nu}\cos\left(\nu\phi/\sqrt{1-\xi_0}\right)}{\left[(1-\xi_0)\varpi^2+z^2\right]^{\nu+3/2}}\label{bz2},
\end{eqnarray}
\end{subequations}
where the normalising factor for the magnetic field is $B_0=\mu_0m/(4\pi R_c^3)$. It is straightforward to verify that the divergence of this field vanishes. For $\nu>-1/2$ (the condition above), all three field components tend to zero in the limit $z\to\pm\infty$, and also for $\varpi\to\infty$. 

As previously discussed, a further requirement is that the pressure, density and current density all vanish as $z\to\pm\infty$. We do not include the expressions here, but it is straightforward to verify that all terms satisfy this requirement when we take $\nu>-1/2$.

It is also straightforward to verify that (\ref{gauss_thm}) is satisfied for the magnetic field given by (\ref{B_varpi}-c),
since the integral over $\phi$ vanishes. In the axisymmetric case, which we can obtain by setting $\nu=0$ (for example), the integral is given (via the divergence theorem) by 
\begin{eqnarray}
\int_{V_c}\left(\bm\nabla{\bm\cdot}\bm{B}\right)\mathrm{d}V=-2\pi R_c^2A_0\int_{-\infty}^{\infty} \left((1-\xi_0)R_c^2+z^2\right)^{-3/2}\mathrm{d}z=-\frac{4\pi A_0}{1-\xi_0},\qquad\label{gauss_thm_special}
\end{eqnarray}
and so it does not vanish. This is due to the fact that, for the special cases considered in this section, $\nu=0$ corresponds to a monopole solution. We therefore restrict attention to solutions depending on all three coordinate variables (when $\mu=-\nu$; the point above does not imply that we cannot have axisymmetric solutions in general). Note, however, that since Laplace's equation is linear, we can superpose linearly independent solutions to construct a new solution and, hence, could superpose one of our new solutions with a known axisymmetric solution of Laplace's equation, such as that of a magnetic dipole aligned with the rotation axis. The only requirement when doing such a superposition is that the constant $\xi_0$ must be the same for each linearly independent solution, for consistency with the coordinate transformation (\ref{transformation}). This point is illustrated in section \ref{sec:fractional}.

\subsection{Example solutions for $\mu=-\nu$}

\label{sec:fractional}

\begin{figure}[htp]
\centering\subfigure[]{\scalebox{0.36}{\includegraphics{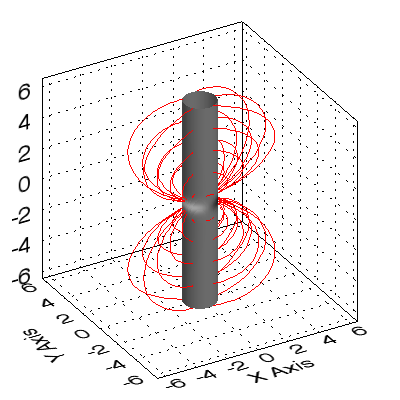}}}
\subfigure[]{\scalebox{0.36}{\includegraphics{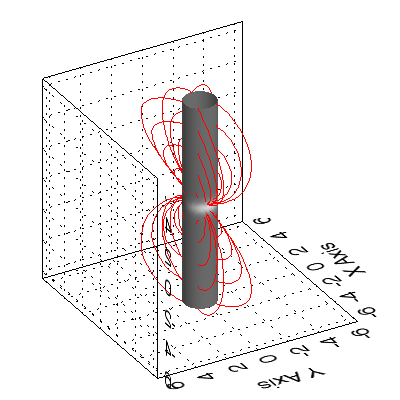}}}
\subfigure[]{\scalebox{0.36}{\includegraphics{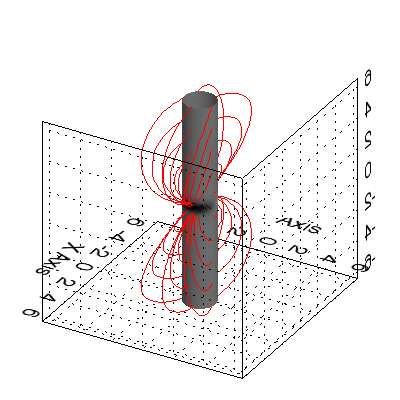}}}
\caption{The field lines and radial magnetic field (shaded) for $\nu=1/2$ and $m=-1$ ($\xi_0=3/4$), shown from three different viewing angles.}\label{fig:NU0PT5_VARYING_N1}
\bigskip
\bigskip
\end{figure}

\begin{figure}[hp]
\centering\subfigure[]{\scalebox{0.36}{\includegraphics{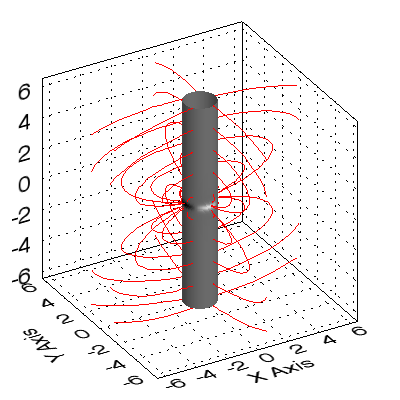}}}
\subfigure[]{\scalebox{0.36}{\includegraphics{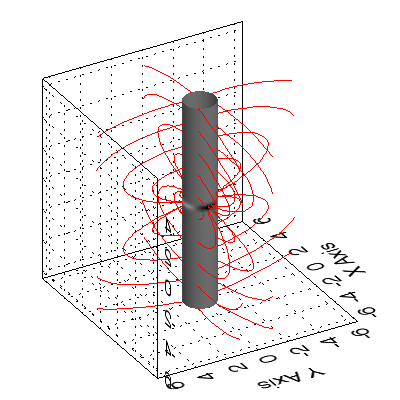}}}
\subfigure[]{\scalebox{0.36}{\includegraphics{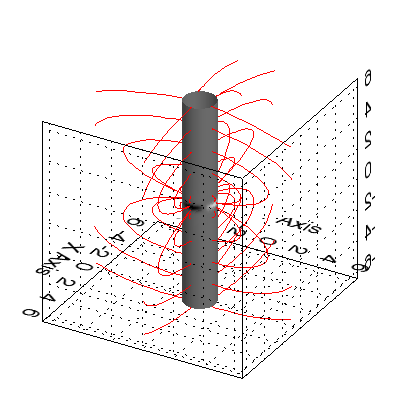}}}
\caption{The field lines and radial magnetic field (shaded) for $\nu=1/2$ and $m=-2$ ($\xi_0=15/16$), shown from three different viewing angles.}\label{fig:NU0PT5_VARYING_N2}
\bigskip
\bigskip
\end{figure}

\begin{figure}[hp]
\centering\subfigure[]{\scalebox{0.36}{\includegraphics{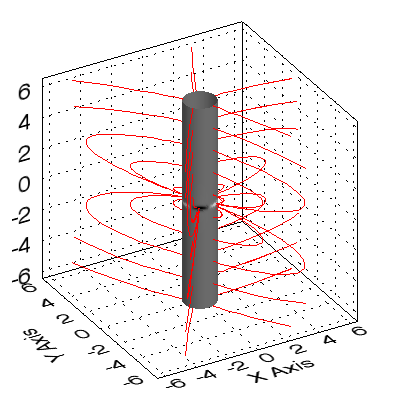}}}
\subfigure[]{\scalebox{0.36}{\includegraphics{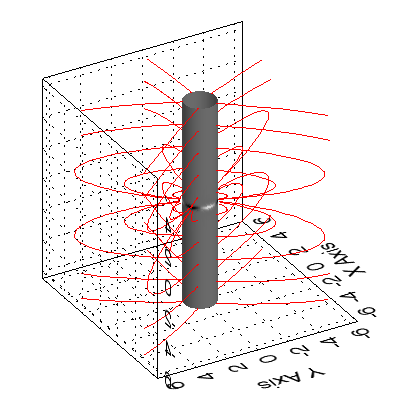}}}
\subfigure[]{\scalebox{0.36}{\includegraphics{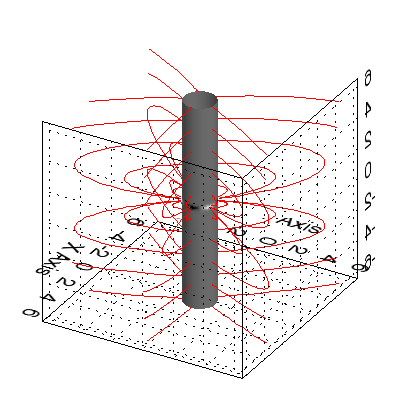}}}
\caption{The field lines and radial magnetic field (shaded) for $\nu=1/2$ and $m=-3$ ($\xi_0=35/36$), shown from three different viewing angles.}\label{fig:NU0PT5_VARYING_N3}
\end{figure}

In this section, we will give some illustrative examples of the solutions with $\mu=-\nu$. For our first example, we take $\nu=1/2$, and hence $\mu=-1/2$. Figures \ref{fig:NU0PT5_VARYING_N1}-\ref{fig:NU0PT5_VARYING_N3} show the magnetic field lines and strength of the radial magnetic field, $B_\varpi$ (shaded on the cylinder), from three different viewing angles, for $m=-1$ (figure \ref{fig:NU0PT5_VARYING_N1}), $m=-2$ (figure \ref{fig:NU0PT5_VARYING_N2}) and $m=-3$ (figure \ref{fig:NU0PT5_VARYING_N3}), i.e. we are varying the dependence of the solution on $\phi$. Since $m$ is related to $\xi_0$ through (\ref{muuuu}) (when $\mu=-\nu$), increasing $\vert m\vert$ with a fixed $\nu$ means that we increase $\xi_0$ - it gets closer and closer to one. This results in a stretching of the field lines, as can be seen from figures  \ref{fig:NU0PT5_VARYING_N1}-\ref{fig:NU0PT5_VARYING_N3}, and the polarity of the radial magnetic field changes more frequently as $\xi_0$ gets closer to one (black and white regions on the cylinder correspond to opposite polarities). Note also that the radial field is strongly confined to the central region of the cylinder in each case (grey regions on the cylinder correspond to those where the radial magnetic field is negligible). 

\begin{figure}[hp]
\centering\subfigure[]{\scalebox{0.36}{\includegraphics{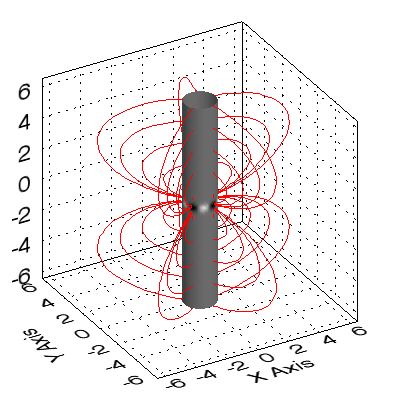}}}
\subfigure[]{\scalebox{0.36}{\includegraphics{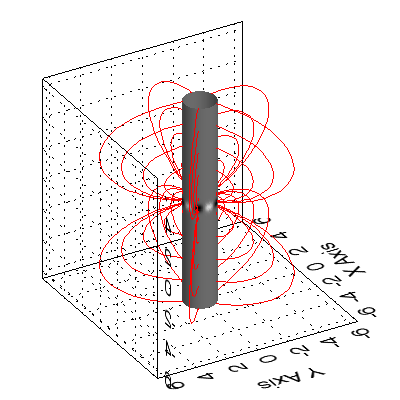}}}
\subfigure[]{\scalebox{0.36}{\includegraphics{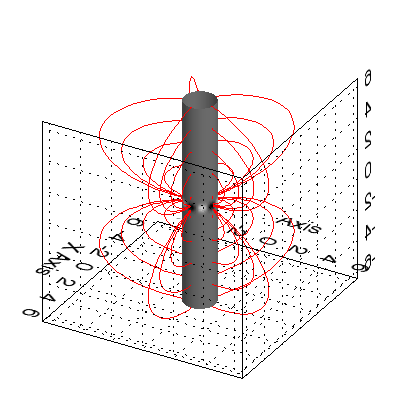}}}
\caption{The field lines and radial magnetic field (shaded) for $\xi_0=3/4$ and $\nu=5/2$, shown from three different viewing angles.}\label{fig:xi0pt75_vary_nu1}
\bigskip
\bigskip
\bigskip
\bigskip
\end{figure}

\begin{figure}[hp]
\centering\subfigure[]{\scalebox{0.36}{\includegraphics{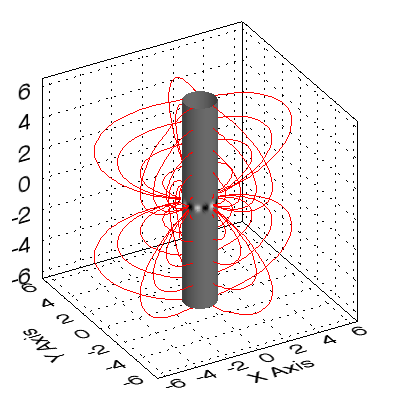}}}
\subfigure[]{\scalebox{0.36}{\includegraphics{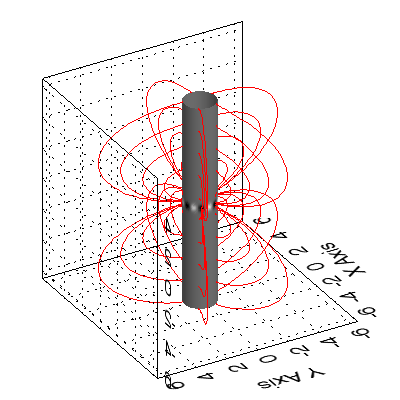}}}
\subfigure[]{\scalebox{0.36}{\includegraphics{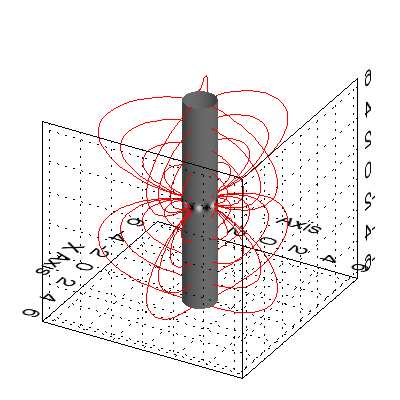}}}
\caption{The field lines and radial magnetic field (shaded) for $\xi_0=3/4$ and $\nu=7/2$, shown from three different viewing angles.}\label{fig:xi0pt75_vary_nu2}
\end{figure}

On the other hand, if we fix $\xi_0$, then varying $m$ will give us different values of $\nu$. We have already considered the case $m=-1$, which gives $\nu=1/2$ when $\xi_0=3/4$ (figure \ref{fig:NU0PT5_VARYING_N1}). Figures \ref{fig:xi0pt75_vary_nu1} and  \ref{fig:xi0pt75_vary_nu2} show the magnetic field lines and radial magnetic field (shaded) for $\xi_0=3/4$, from three different viewing angles, for $\nu=5/2$, i.e. $m=-5$ (figure \ref{fig:xi0pt75_vary_nu1}) and $\nu=7/2$, i.e. $m=-7$ (figure \ref{fig:xi0pt75_vary_nu2}). It can be seen that the radial stretching of the field lines is roughly the same in each case due to the solutions having the same value of $\xi_0$, but in this case increasing $\nu$ corresponds to increasing $\vert m\vert$, i.e. changing the $\phi$ dependence of the solutions, which can be seen from the polarity of the radial magnetic field (plotted on the surface of the cylinder); there are more frequent changes in the polarity for higher values of $\nu$.

\begin{figure}[htp]
\centering\subfigure[]{\scalebox{0.23}{\includegraphics{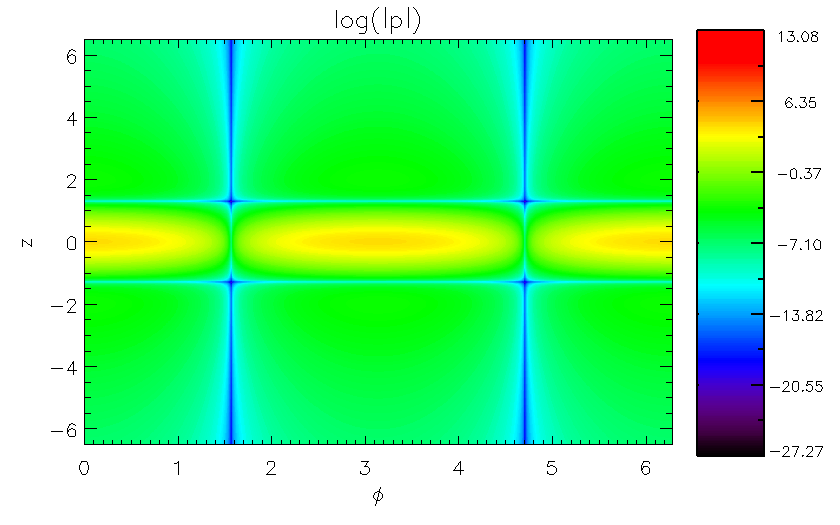}}}\subfigure[]{\scalebox{0.23}{\includegraphics{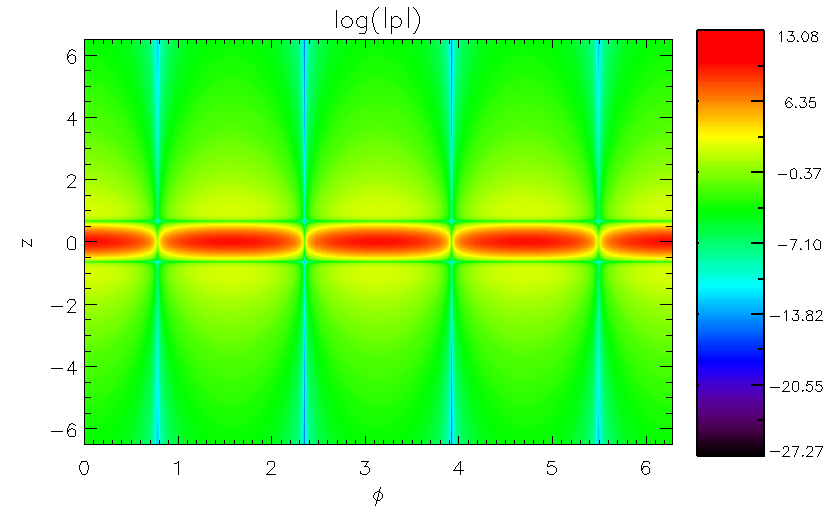}}}\\
\subfigure[]{\scalebox{0.23}{\includegraphics{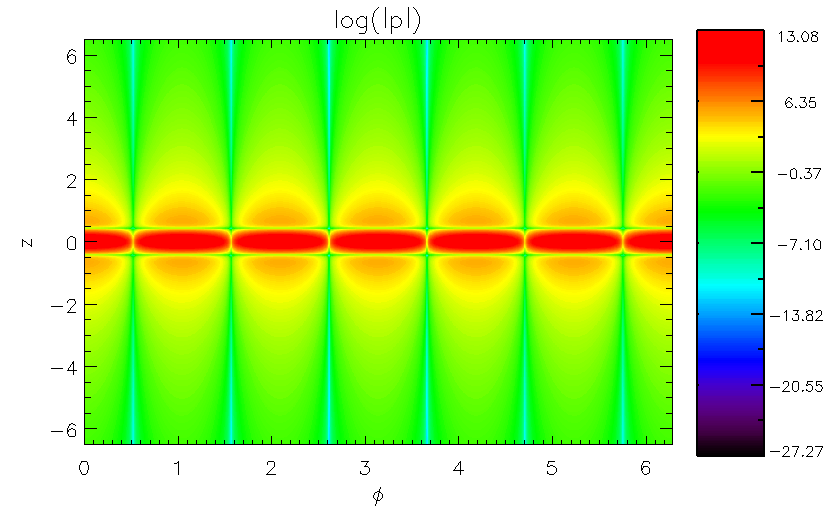}}}
\caption{Plots showing the pressure variation for $\nu=1/2$ and (a) $\xi_0=3/4$, (b) $\xi_0=15/16$, (c) $\xi_0=35/36$, at $\varpi=1.5$. (colour online)}\label{fig:frac_p}
\end{figure}

\begin{figure}[htp]
\centering\subfigure[]{\scalebox{0.23}{\includegraphics{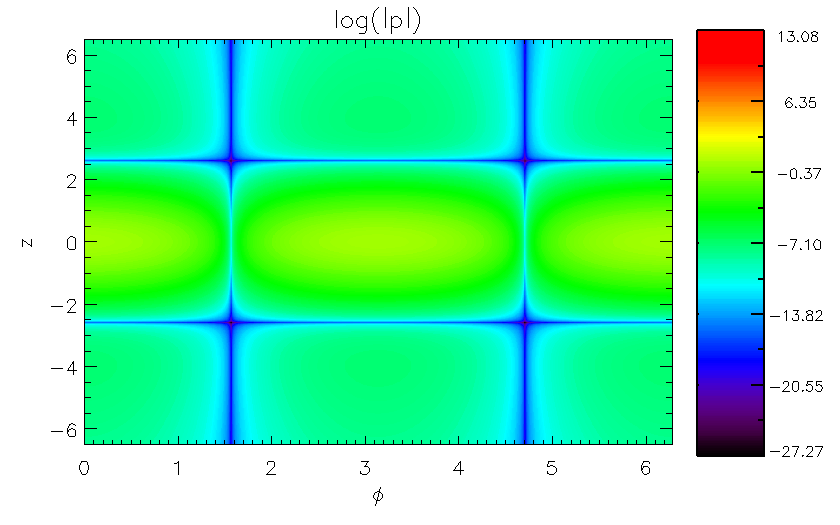}}}\subfigure[]{\scalebox{0.23}{\includegraphics{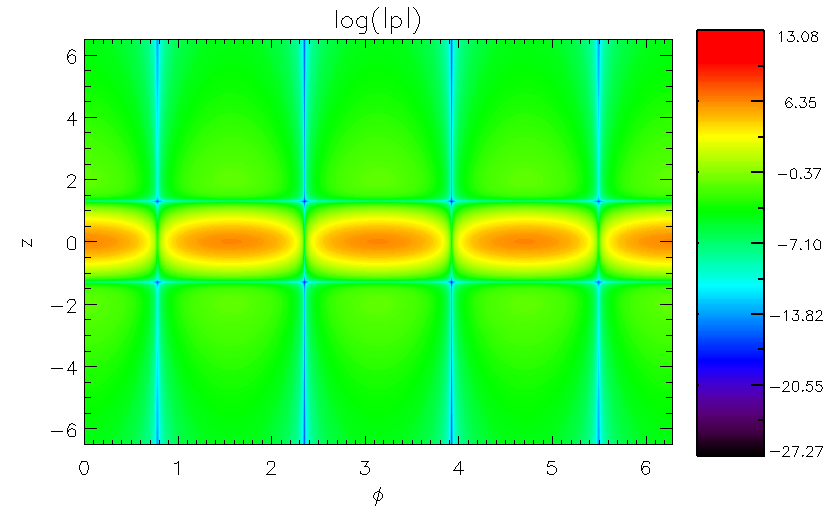}}}\\
\subfigure[]{\scalebox{0.23}{\includegraphics{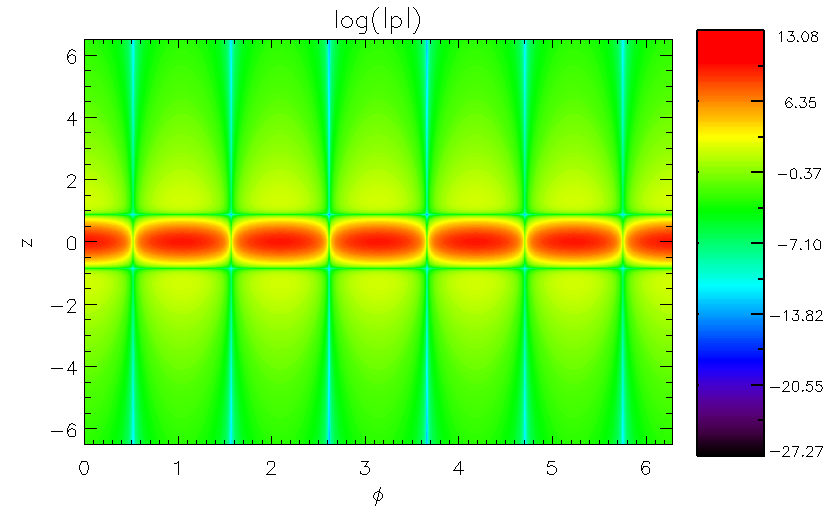}}}
\caption{Plots showing the pressure variation for $\nu=1/2$ and (a) $\xi_0=3/4$, (b) $\xi_0=15/16$, (c) $\xi_0=35/36$, at $\varpi=3.0$. (colour online)}\label{fig:frac_p2}
\end{figure}

For a complete MHS equilibrium solution, we also need to find the pressure and density, and specify the temperature through an equation of state. To illustrate the effect of changing $\xi_0$, we will now show plots of the pressure and density variation for the parameters in figures \ref{fig:NU0PT5_VARYING_N1}-\ref{fig:NU0PT5_VARYING_N3} (obtained by subtracting the background terms). Figures \ref{fig:frac_p} and \ref{fig:frac_p2} show contours of the natural logarithm of the pressure variation at $\varpi=1.5$ and $\varpi=3.0$, respectively. 

\begin{figure}[bp]
\centering\subfigure[]{\scalebox{0.23}{\includegraphics{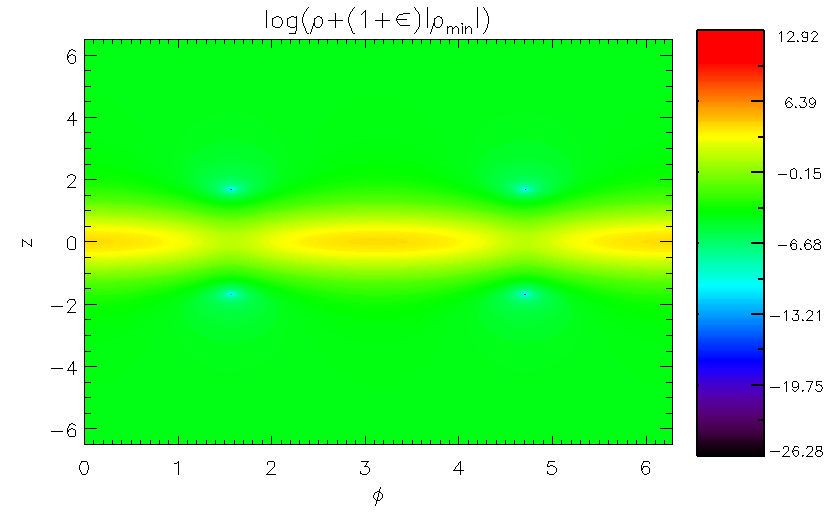}}}\subfigure[]{\scalebox{0.23}{\includegraphics{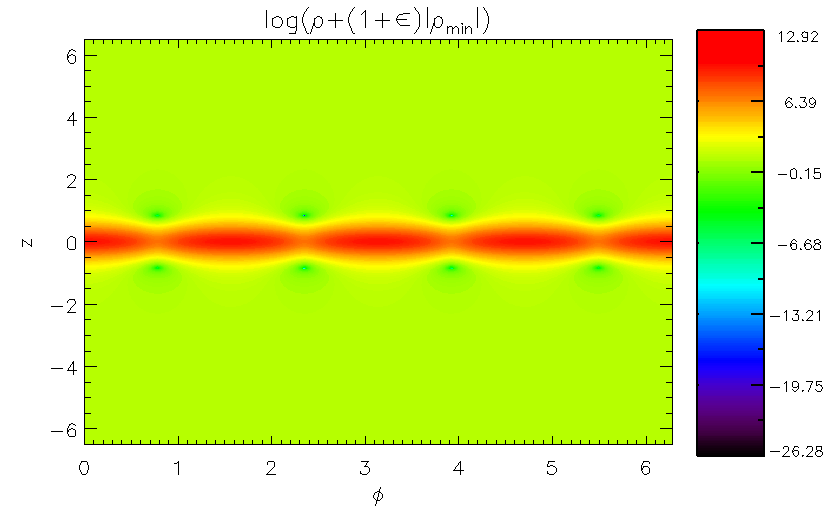}}}\\
\subfigure[]{\scalebox{0.23}{\includegraphics{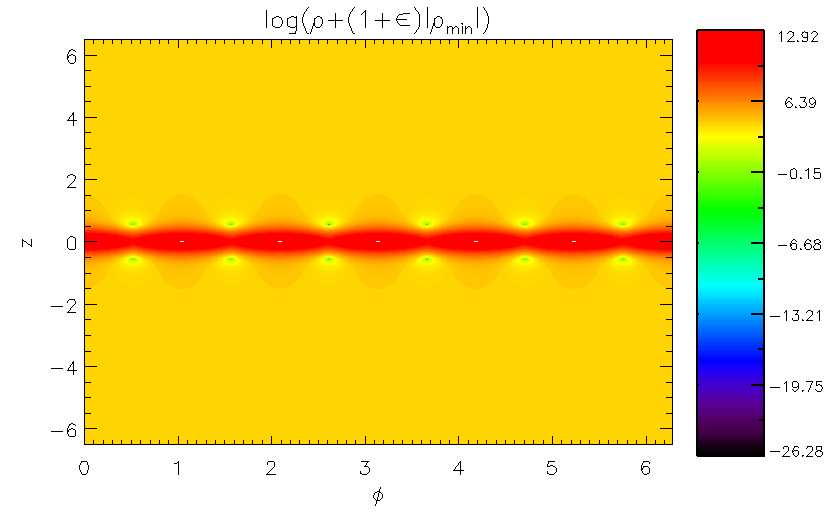}}}
\caption{Plots showing the density variation for $\nu=1/2$ and (a) $\xi_0=3/4$, (b) $\xi_0=15/16$, (c) $\xi_0=35/36$, at $\varpi=1.5$. (colour online)}\label{fig:frac_d}
\end{figure}

\begin{figure}[htp]
\centering\subfigure[]{\scalebox{0.23}{\includegraphics{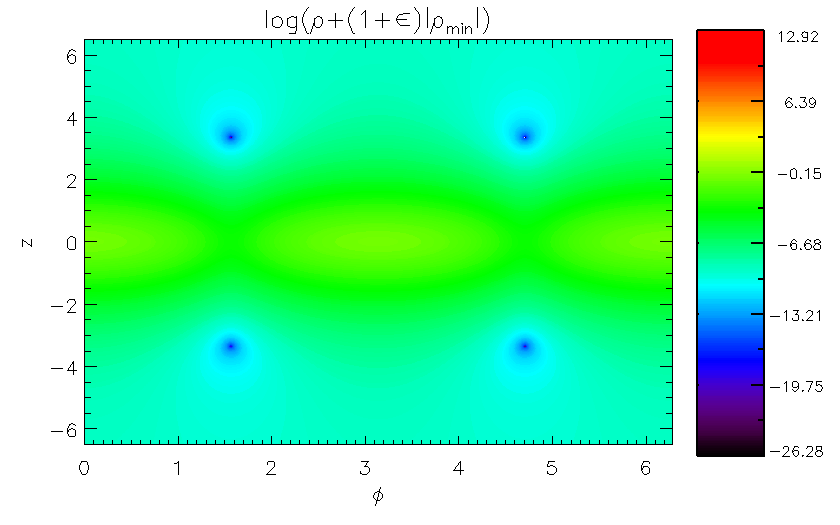}}}\subfigure[]{\scalebox{0.23}{\includegraphics{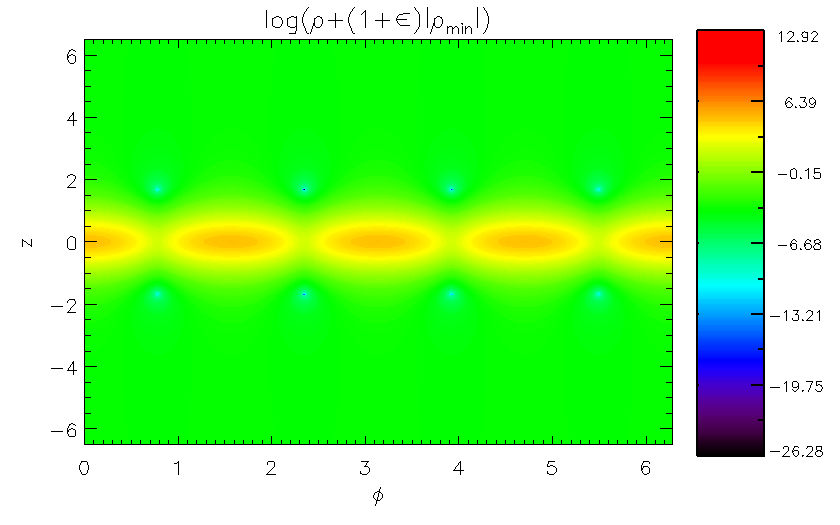}}}\\
\subfigure[]{\scalebox{0.23}{\includegraphics{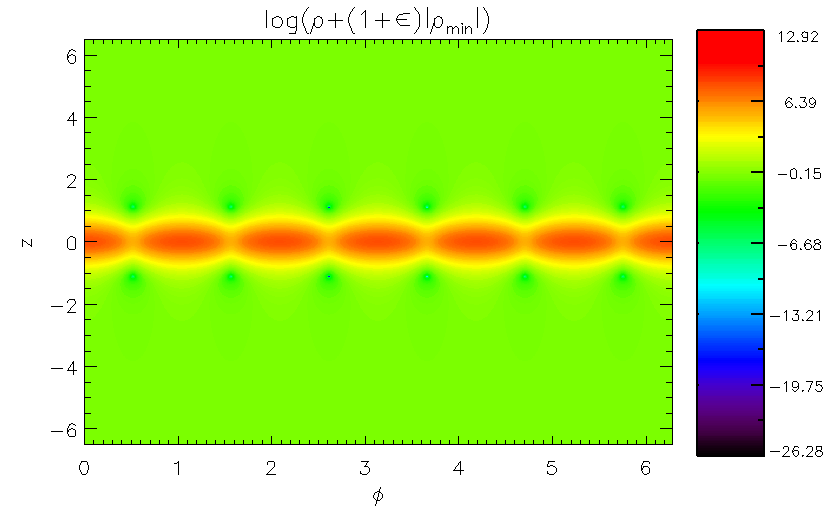}}}
\caption{Plots showing the density variation for $\nu=1/2$ and (a) $\xi_0=3/4$, (b) $\xi_0=15/16$, (c) $\xi_0=35/36$, at $\varpi=3.0$. (colour online)}\label{fig:frac_d2}
\end{figure}

From (\ref{pressure_diff}), it is clear that the pressure variation is always negative for a constant $\xi_0$ and, hence, we plot the natural logarithm of its modulus. This means that the pressure will always be less than the chosen background pressure. Figures \ref{fig:frac_d} and \ref{fig:frac_d2} show contours of the natural logarithm of the density variation at $\varpi=1.5$ and $\varpi=3.0$, respectively. The density variation can be both positive and negative, meaning that the density can be either enhanced or reduced compared to the chosen background density. For plotting the density variation, therefore, we add a constant $(1+\epsilon)\vert\rho_{min}\vert$ before taking the natural logarithm, where $\epsilon$ is a chosen ``small"


\FloatBarrier
\noindent parameter and $\rho_{min}$ is the minimum value of the density in the domain. In doing this, we are shifting the values of the function on the contours, but still capturing the shape of the contours. 

From figures \ref{fig:frac_p}-\ref{fig:frac_d2}, the periodicity in the density and pressure can be seen, and it is clear that, as $\xi_0$ increases (becomes closer to 1), there are a greater number of oscillations in the pressure and density. This is consistent with there being a greater number of polarity changes in the radial magnetic field as $\xi_0$ is increased. It can also be seen that both the pressure and density decrease with increasing radial coordinate in each case. The normalising factors for the pressure and density are $B_0^2/(2\mu_0)=\mu_0m_d^2/(32\pi^2R_c^6)$ and $\rho_0=\mu_0m_d^2/(32\pi^2\varOmega^2R_c^8)$, respectively.

In this section, we have given some simple example solutions for the special case $\mu=-\nu$. We have chosen non-integer values of $\nu$ in each example, and so the solutions shown resemble fractional multipoles \citep{Engheta-1996,Debnath-2003}). For integer values of $\nu$, however, we note that the magnetic field, pressure and density all have a structure which is qualitatively similar to that of the fractional multipole-type solutions and so, in that sense, there is nothing ``special" about the fractional multipole solutions compared to the integer multipole ones (for this special class of solutions).

\subsection{Example: superposition of solutions}

As we have previously discussed, Laplace's equation is linear, and so we can superpose linearly independent solutions in any way we like to construct a new solution. We will briefly illustrate this point in the present section. 

We could, e.g., superpose the solutions in figures \ref{fig:NU0PT5_VARYING_N1}, \ref{fig:xi0pt75_vary_nu1} and \ref{fig:xi0pt75_vary_nu2} to create a new solution. We will not demonstrate this, however, since the structure of the field lines is relatively similar in each case. We could also superpose one (or as many as we like) of the solutions with, e.g. an axisymmetric dipole solution, or a tilted/shifted dipole solution. \citet{Neukirch-2009b} considered a dipole solution of the form
\begin{subequations}
\begin{equation}
U(\varpi,\phi,z)\,=\,-\,A_d\frac{f(\varpi,\phi,z)}{g(\varpi,\phi,z)}\,,
\end{equation}
where
\begin{align}
  f(\varpi,\phi,z) \, =\, &\,\sin\varTheta\cos\varPsi\left(\varpi\cos\left(\frac{\phi}{\sqrt{1-\xi_0}}\right)-\varpi_d\right)\nonumber\\
  &\,\hskip 5mm+\sin\varTheta\sin\varPsi\sin\left(\frac{\phi}{\sqrt{1-\xi_0}}\right)\varpi+\frac{z\cos\varTheta}{\sqrt{1-\xi_0}}\,,\\
g(\varpi,\phi,z)\,=\,&\,{\left[\varpi^2-2\varpi_d\varpi\cos\left(\frac{\phi}{\sqrt{1-\xi_0}}\right)+\varpi_d^2+\frac{z^2}{(1-\xi_0)}\right]^{3/2}},
\end{align}
\end{subequations}
where $A_d$ is a constant and the ``pseudo-potential" $U$ is normalised by $\mu_0m_d/(4\pi R_c^{2})$; the angles $\varTheta$ and $\varPsi$ are the polar and azimuthal angles for the dipole direction, and $\varpi_d$ is the displacement of the dipole along the $x$-axis. Note that the dipole solution above with $\varTheta=0$ and $\varpi_d=0$ can be obtained from the solution in (\ref{laplace_sol_cyl}) by normalising in the same way as above, and setting $\mu=0$, $\nu=1$, $A_{01}=-(1-\xi_0)A_d/2$, $C_{01}=1$ and $D_{01}=-1$; it is an example of an integer multipole case.

\begin{figure}[htp]
\bigskip
\bigskip
\centering\scalebox{0.36}{\includegraphics{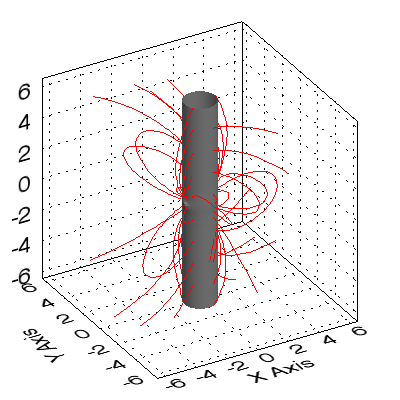}}
\scalebox{0.36}{\includegraphics{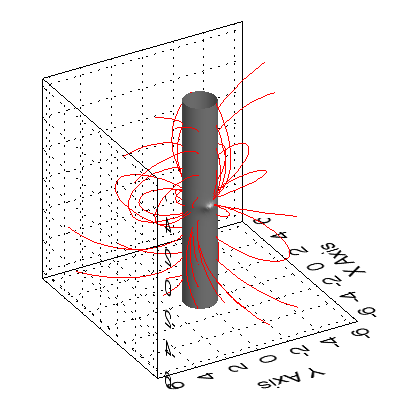}}
\scalebox{0.36}{\includegraphics{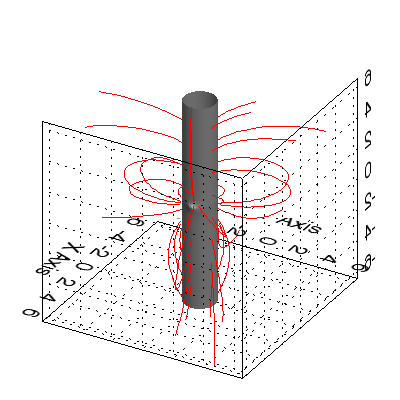}}
\caption{The dipole solution considered by \citet{Neukirch-2009b}, which is centred at $x=\varpi_d=0.5$, and tilted by the angles $\varTheta=\varPsi=\pi/4$ (shown from three different viewing angles).}\label{fig:n2009_dipole}
\bigskip
\bigskip
\end{figure}

\begin{figure}[htp]
\centering\scalebox{0.36}{\includegraphics{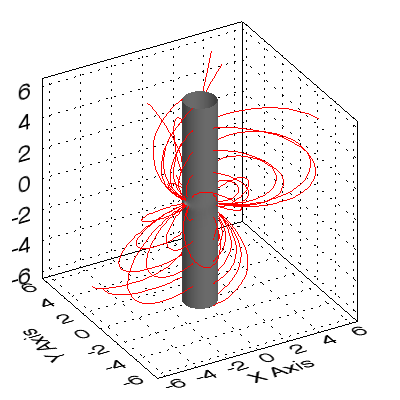}}
\scalebox{0.36}{\includegraphics{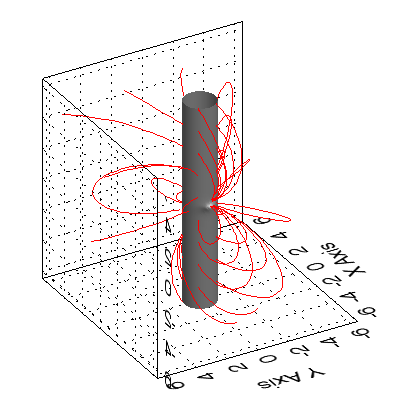}}
\scalebox{0.36}{\includegraphics{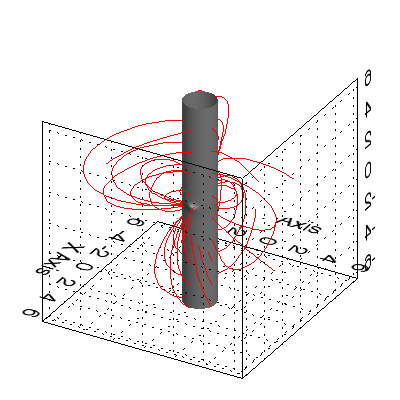}}
\caption{The superposition of the titled dipole solution from figure \ref{fig:n2009_dipole} with the fractional multipole solution shown in Fig 2a, for $\xi_0=3/4$ and $\nu=1/2$, for $A_\nu/A_d=0.1$, shown from three different viewing angles.}\label{fig:superposition1}
\end{figure}

\begin{figure}[htp]
\centering\scalebox{0.36}{\includegraphics{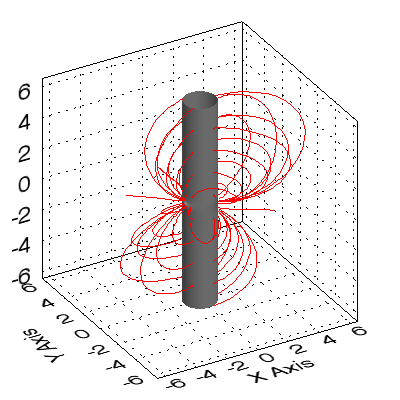}}
\scalebox{0.36}{\includegraphics{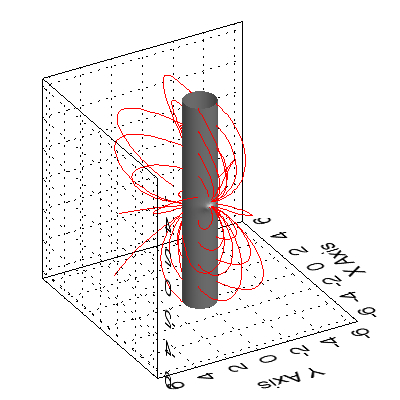}}
\scalebox{0.36}{\includegraphics{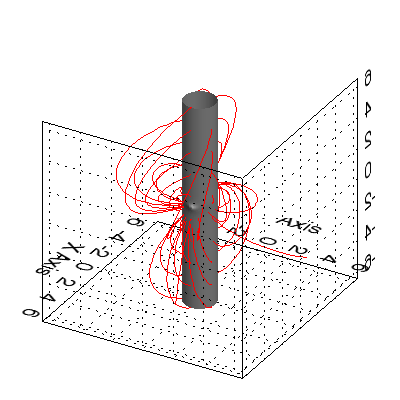}}
\caption{The superposition of the titled dipole solution from figure \ref{fig:n2009_dipole} with the fractional multipole solution shown in Fig 2a, for $\xi_0=3/4$ and $\nu=1/2$, for $A_\nu/A_d=0.2$, shown from three different viewing angles.}\label{fig:superposition2}
\bigskip
\bigskip
\bigskip
\bigskip
\end{figure}

\begin{figure}[htp]
\centering\scalebox{0.36}{\includegraphics{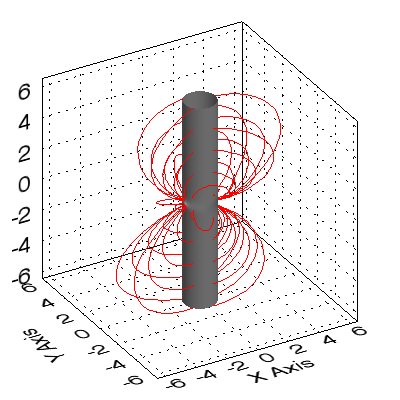}}
\scalebox{0.36}{\includegraphics{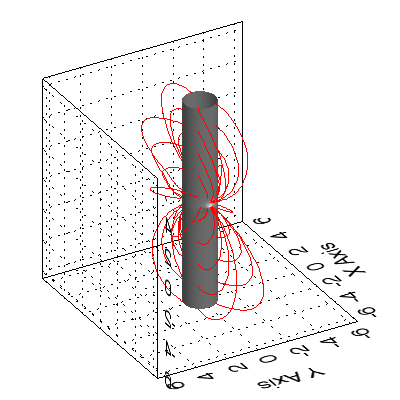}}
\scalebox{0.36}{\includegraphics{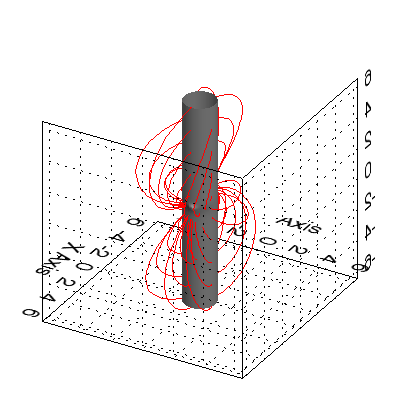}}
\caption{The superposition of the titled dipole solution from figure \ref{fig:n2009_dipole} with the fractional multipole solution shown in Fig 2a, for $\xi_0=3/4$ and $\nu=1/2$, for $A_\nu/A_d=0.4$, shown from three different viewing angles.}\label{fig:superposition3}
\end{figure}

For illustration, the parameters used by \citet{Neukirch-2009b} were $\xi_0=3/4$, $\varTheta=\psi=\pi/4$, $\varpi_d=0.5$. This solution is shown in figure \ref{fig:n2009_dipole}. Figures \ref{fig:superposition1}-\ref{fig:superposition3} show the result of superposing this dipole solution with the fractional multipole solution shown in figure \ref{fig:NU0PT5_VARYING_N1}, for $\xi_0=3/4$, $\nu=1/2$, and three different values of the ratio $A_v/A_d$, which represents the relative size of the two solutions superposed.

\begin{figure}[htp]
\centering\subfigure[]{\scalebox{0.24}{\includegraphics[trim=20 10 0 5,clip]{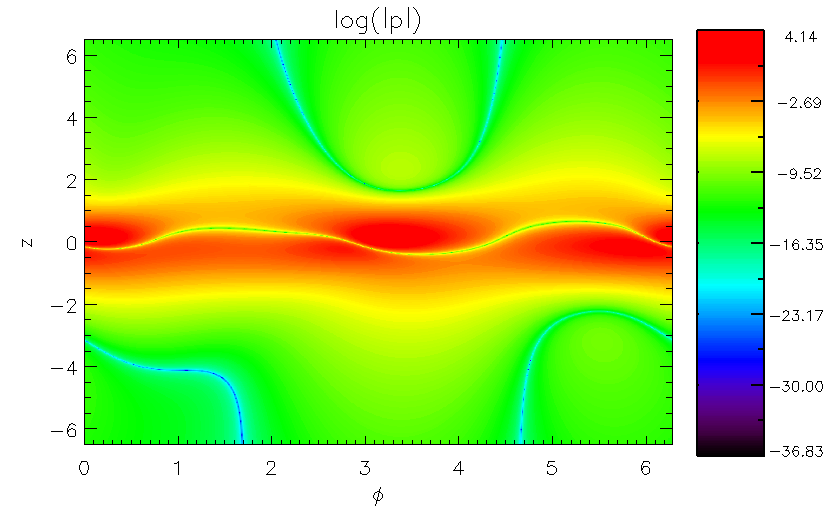}}}\subfigure[]{\scalebox{0.24}{\includegraphics[trim=20 10 0 5,clip]{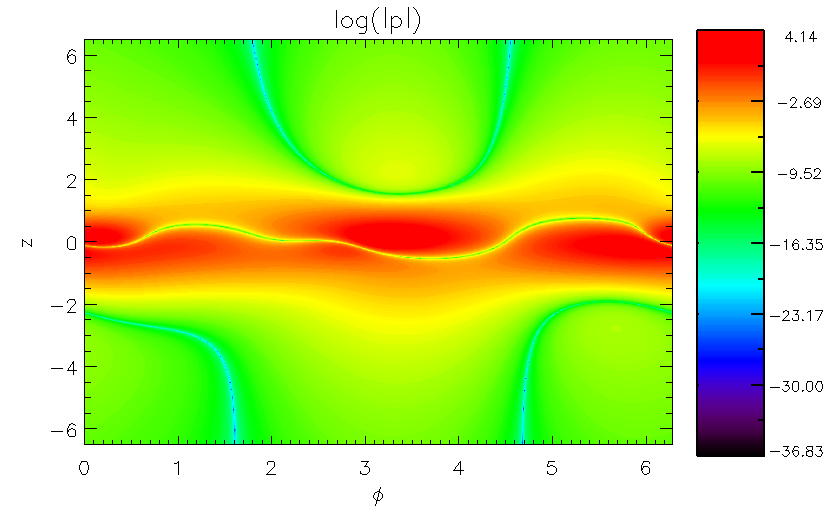}}}\\
\subfigure[]{\scalebox{0.24}{\includegraphics[trim=20 10 0 5,clip]{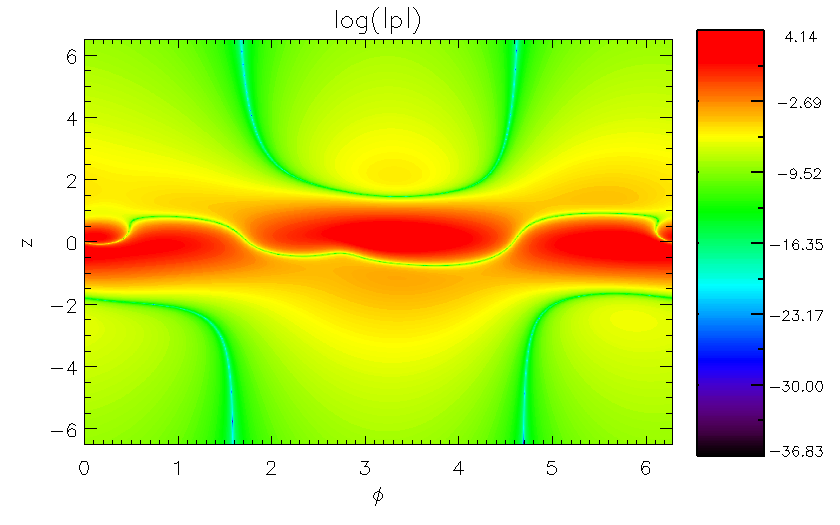}}}
\caption{Plots of the pressure variation for the superposition of solutions shown in figures \ref{fig:superposition1}-\ref{fig:superposition3} with (a) $A_\nu/A_d=0.1$, (b) $A_\nu/A_d=0.2$, (c) $A_\nu/A_d=0.4$, at $\varpi=1.5$. (colour online)}\label{fig:super_p}
\end{figure}

\begin{figure}[htp]
\centering\subfigure[]{\scalebox{0.24}{\includegraphics[trim=20 10 0 5,clip]{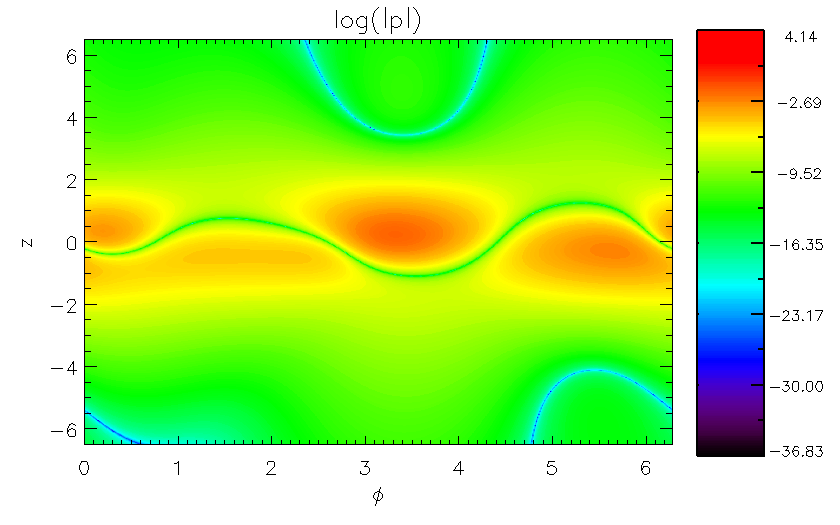}}}\subfigure[]{\scalebox{0.24}{\includegraphics[trim=20 10 0 5,clip]{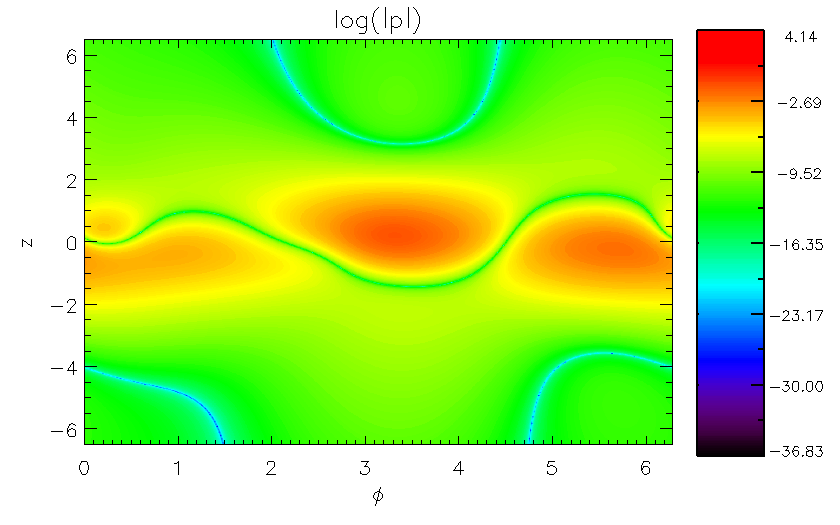}}}\\
\subfigure[]{\scalebox{0.24}{\includegraphics[trim=20 10 0 5,clip]{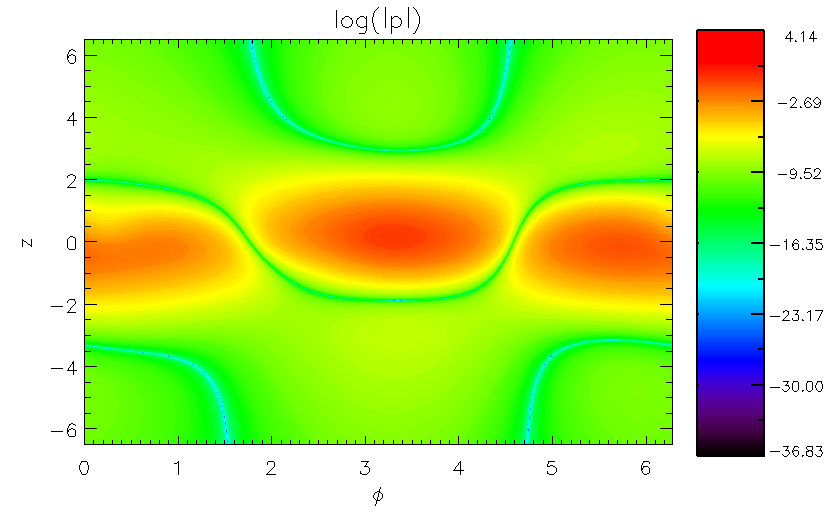}}}
\caption{Plots of the pressure variation for the superposition of solutions shown in figures \ref{fig:superposition1}-\ref{fig:superposition3} with (a) $A_\nu/A_d=0.1$, (b) $A_\nu/A_d=0.2$, (c) $A_\nu/A_d=0.4$, at $\varpi=3.0$. (colour online)}\label{fig:super_p2}
\end{figure}

\begin{figure}[htp]
\centering\subfigure[]{\scalebox{0.24}{\includegraphics[trim=20 10 0 5,clip]{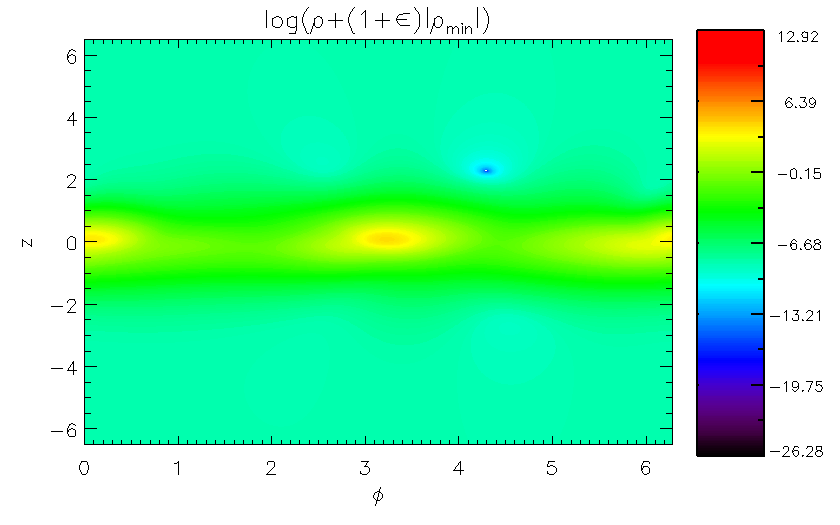}}}\subfigure[]{\scalebox{0.24}{\includegraphics[trim=20 10 0 5,clip]{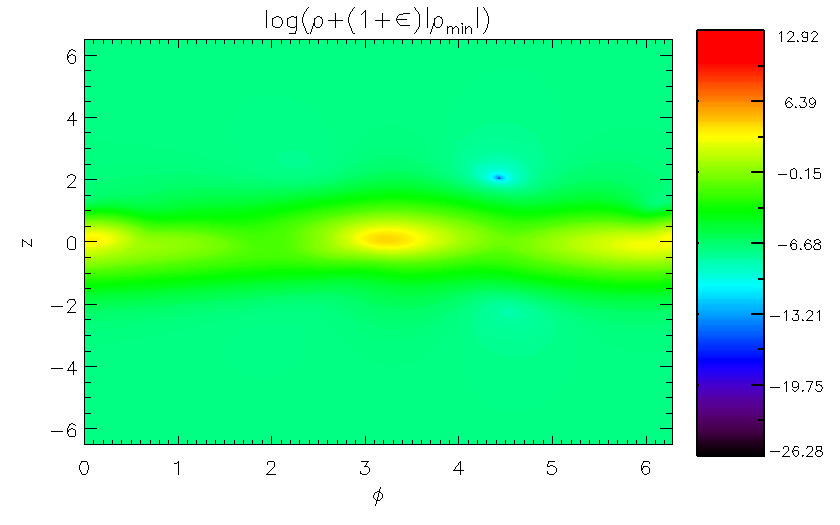}}}\\
\subfigure[]{\scalebox{0.24}{\includegraphics[trim=20 10 0 5,clip]{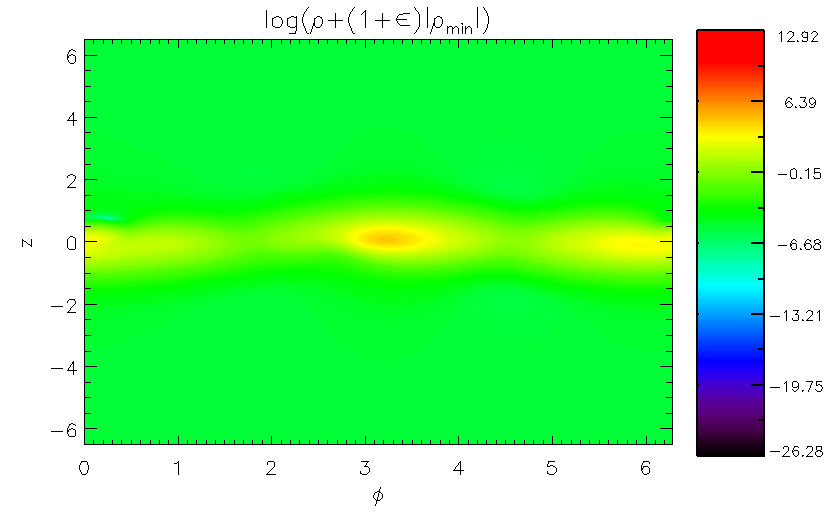}}}
\caption{Plots of the density variation for the superposition of solutions shown in figures \ref{fig:superposition1}-\ref{fig:superposition3} with (a) $A_\nu/A_d=0.1$, (b) $A_\nu/A_d=0.2$, (c) $A_\nu/A_d=0.4$, at $\varpi=1.5$. (colour online)}\label{fig:super_d}
\end{figure}

\begin{figure}[htp]
\centering\subfigure[]{\scalebox{0.24}{\includegraphics[trim=20 10 0 5,clip]{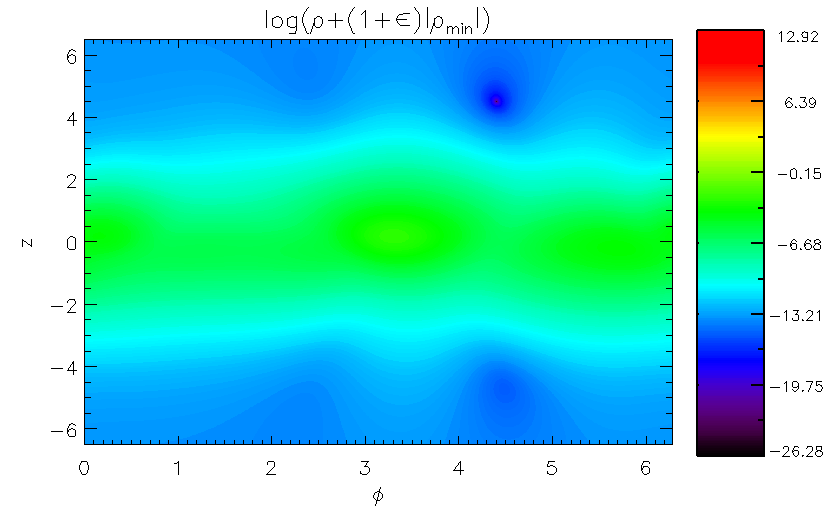}}}\subfigure[]{\scalebox{0.24}{\includegraphics[trim=20 10 0 5,clip]{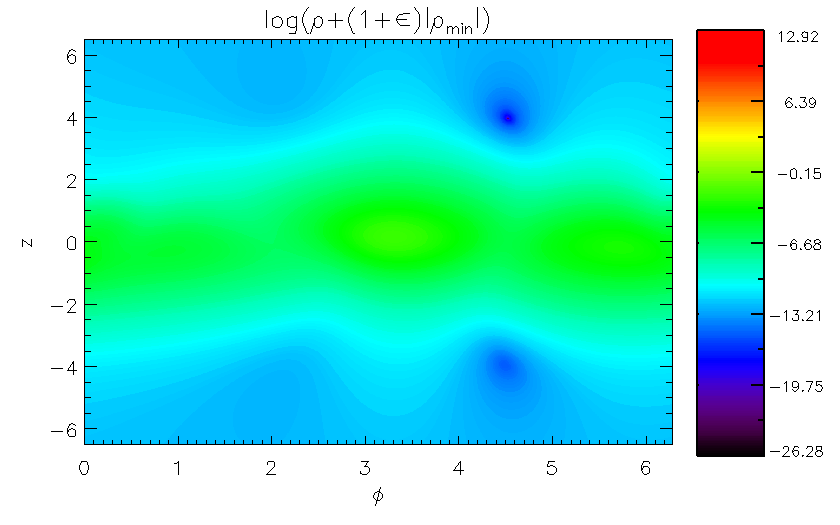}}}\\
\subfigure[]{\scalebox{0.24}{\includegraphics[trim=20 10 0 5,clip]{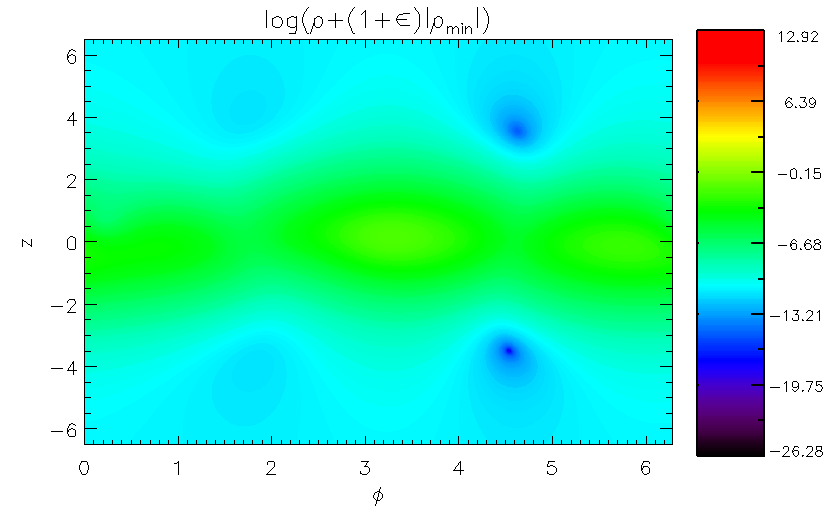}}}
\caption{Plots of the density variation for the superposition of solutions shown in figures \ref{fig:superposition1}-\ref{fig:superposition3} with (a) $A_\nu/A_d=0.1$, (b) $A_\nu/A_d=0.2$, (c) $A_\nu/A_d=0.4$, at $\varpi=3.0$. (colour online)}\label{fig:super_d2}
\end{figure}

Figures \ref{fig:super_p} and \ref{fig:super_p2} show contours of the natural logarithm of the pressure variation for the superposed solution at $\varpi=1.5$ and $\varpi=3.0$, respectively, for the values of the ratio $A_\nu/A_d$ taken in figures \ref{fig:superposition1}-\ref{fig:superposition3}. Again, from (\ref{pressure_diff}), we have that the pressure variation is always negative for a constant $\xi_0$, and so we plot the natural logarithm of its modulus. Figures \ref{fig:super_d} and \ref{fig:super_d2} show contours of the natural logarithm of the density variation, again for $\varpi=1.5$ and $\varpi=3.0$ and the three different values of the ratio $A_\nu/A_d$. As before, we have that the density variation can be positive or negative, and so we add a constant $(1+\epsilon)\vert\rho_{min}\vert$ before taking the natural logarithm, where $\epsilon$ is a chosen ``small" parameter and $\rho_{min}$ is the minimum value of the density in the domain.

It can be seen that, in the three cases, there is a different periodicity in $\phi$ from the examples previously discussed, since the superposed dipole solution has a periodicity of $2\phi$. For a given value of the ratio $A_\nu/A_d$, it can be seen that both the pressure and density variations decrease as the radial coordinate increases. For a fixed value of the radial coordinate, increasing the ratio $A_\nu/A_d$ results in a noticeable alteration of the pressure and density structure in each case, since the fractional multipole part has a different structure than the dipole component, which becomes more pronounced as the relative amplitude increases.\\

\FloatBarrier
\section{Summary and conclusions}

\label{sec:summary}

In this paper, we have presented further three-dimensional analytical solutions of the MHS equilibrium problem in the co-rotating frame of reference outside a magnetised rigidly rotating cylindrical body, for the case with only centrifugal force (neglecting gravity), which extends previous work by \citet{Neukirch-2009b}. We have used the method developed by Low \citep[e.g.][]{low-1985,low-1991,low-1992,low-1993a,low-1993b} and a transformation method used by \citet{Neukirch-2009b}. The governing equation for the ``pseudo-potential" (from which the components of the magnetic field can be derived) can be transformed into Laplace's equation, for which many simple solutions are known. 

We have shown how, through the use of ``pseudo-spherical" coordinates, we can find further separable cylindrical solutions. These solutions contain associated Legendre functions, which in general can have any order and/or degree. This, for example, gives rise to fractional multipole solutions \citep[see, e.g.,][]{Engheta-1996,Debnath-2003} which, to the best of our knowledge, have not been discussed before in the context of MHS equilibria. The further solutions also have a less restrictive dependence on the azimuthal coordinate, $\phi$, than those discussed by \citet{Neukirch-2009b} - they can depend on $\phi$, and so no longer have a lowest order dependence of $2\phi$.

We have presented some simple example solutions for the case when the degree and order of the associated Legendre functions sum to zero; in this case a simple closed form can be used, which allows us to ensure (in a straightforward way) that the solution does not contain any singularities. It should be noted, however, that for these example cases we restrict attention to solutions depending on all three coordinate variables, since the axisymmetric solutions (i.e. those which are independent of $\phi$) do not satisfy the requirement that the integral of $\bm\nabla{\bm\cdot}\bm{B}$ vanishes over the cylindrical volume.

Since Laplace's equation is linear, we can also superpose solutions in any way we like in order to construct a new solution. We demonstrated this by showing an example solution constructed by superposing one of our new solutions with a dipole solution that was previously discussed by \citet{Neukirch-2009b}.

It is important to bear in mind that we have considered an idealised case of an infinitely long cylinder, which was done in order to make analytical progress. The boundary conditions will be different for each case, and simply depend on what the particular solution looks like on the boundary of the cylinder. We have shown that superposition of solutions is possible, and an interesting point to explore would be whether the method we have used can capture every possibility.

For applications to, e.g., stellar physics, it would be more useful to consider MHS equilibrium solutions outside a rigidly rotating spherical body. We have attempted to do this using the methods presented in this paper. A problem, however, is that for our choice $\kappa(V)=\xi_0/\left(\bm\nabla V\right)^2$, the density becomes singular on the rotation axis.  We could avoid this problem by considering only values of $\bar\theta$ in the range $0<\bar\theta_{min}\le\bar\theta\le\bar\theta_{max}<\pi$ (i.e. excluding points on the rotation axis), which would require us to specify boundary conditions for either $U$ or $\bm{B}$ at $\bar\theta_{min}$ and $\bar\theta_{max}$. Alternatively, we could take a different choice of $\kappa(V)$ to make the expression for the density non-singular, but this would force us to use a different method, since the transformation method discussed here would no longer work. For example, we could use a numerical method similar to that discussed by  \citet{Al-Salti-2010a} and \citet{Al-Salti-2010b}.\\

\section*{Acknowledgements}%

\noindent The authors would like to thank the anonymous referees, whose comments have helped to improve this manuscript. This work was supported by the Science and Technology Facilities Council under grants ST/K000950/1 and ST/N000609/1.

\section*{Disclosure statement}

\noindent The authors have no potential conflict of interest to report.

\newcommand{\noopsort}[1]{} \newcommand{\printfirst}[2]{#1}
  \newcommand{\singleletter}[1]{#1} \newcommand{\switchargs}[2]{#2#1}

\appendices{
\section{Choosing the parameters $\mu$ and $\nu$}
In section \ref{sec:cyl_new}, we discussed how the Legendre functions in the solution (\ref{laplace_sol_cyl}) can be singular when their argument equals $\pm1$, corresponding to $z\to\pm\infty$ (and $\bar{\varpi}=0$, but we do not need to consider this limit, as discussed in the main text). This places a restriction on the choices we can take for the parameters $\mu$ and $\nu$, which represent the order and degree of the Legendre functions, respectively. These parameters must be chosen in such a way that the ``pseudo-potential" $U(\bar{\varpi},\bar{\phi},z)$ and all of the terms in the force balance equation (\ref{force-ampere-divB}a)
vanish as $z\to\pm\infty$  (i.e. on the upper and lower boundaries of the cylindrical domain).

In the discussion below, we restrict attention to real-valued solutions of the Legendre equation, which are also known as Ferrers' functions \citep{NIST:DLMF}. As the argument of either of the Legendre functions in (\ref{laplace_sol_cyl}) approaches 1 from below, i.e. as $z\to\infty$ for the first function, and $z\to-\infty$ for the second function, they have the asymptotic form \citep{NIST:DLMF}
\begin{equation}
{\mathrm P}_{\nu}^{\mu}(x)\sim\frac{1}{\Gamma(1-\mu)}\left(\frac{2}{1-x}\right)^{\mu/2},\label{asymp}
\end{equation}
where $\Gamma$ is the gamma function. This formula applies if $\mu$ is not a positive integer (the gamma function has poles when its argument is equal to zero or a negative integer). We will first consider part of the solution in (\ref{laplace_sol_cyl}), given by
\begin{equation}
 U_1(\bar{\varpi},z)=\left(\bar{\varpi}^2+z^2\right)^{-(\nu+1)/2}{\mathrm P}_\nu^\mu\left(\frac{z}{\sqrt{\bar{\varpi}^2+z^2}}\right),\label{u2}
\end{equation}
where we do not include the $\bar{\phi}$ dependence, or the constants $A_{\mu\nu}$ and $C_{\mu\nu}$, since they play no role in the discussion. We can show that we require $\mu<\nu+1$ such that $U_1\to0$ when $z\to\infty$, i.e. when the argument of the Legendre function in (\ref{u2}) approaches 1 from below. In this limit, we have (using (\ref{asymp}))
\begin{subequations}
\begin{equation}
U_1(\bar{\varpi},z)\propto \left(1+\frac{\bar\varpi^2}{z^2}\right)^{\mu/4-(\nu+1)/2}G(\bar\varpi,z),\label{u1_prop}
\end{equation}
where
\begin{equation}
G(\bar\varpi,z)=z^{-(\nu+1)}\left(\sqrt{1+\frac{\bar\varpi^2}{z^2}}-1\right)^{-\mu/2}.
\end{equation}
\end{subequations}
For $z\to\infty$, the function multiplying $G(\bar\varpi,z)$ in (\ref{u1_prop}) tends to 1, and so we need to calculate the range of values of $\mu$ and $\nu$ such that $G\to0$ for $z\to\infty$, to determine when $U_1\to0$ in this limit. We only need to consider the case when $\mu>0$, since, if $\mu<0$, the limit will vanish regardless of the values of $\mu$ and $\nu$ (provided $\nu>-1$). Letting $X=z^{-2}$, we have that
\begin{eqnarray}
\lim_{z\to\infty}G(\bar\varpi,z)&=&\lim_{X\to0}\left[\frac{X^{(\nu+1)/2}}{\left(\sqrt{1+\bar\varpi^2X}-1\right)^{\mu/2}}\right].
\end{eqnarray}
By applying L'H\^{o}pital's rule several times, it becomes apparent that we must repeat the process $n$ times, until $\mu/2-n>0$, i.e. so that we do not have zero in the denominator when taking the limit. At the same time, we require that $(\nu+1)/2-n>0$. Combining the conditions on $\mu$ and $\nu$ gives
\begin{equation}
{(\nu+1)}/{2}\,>\, n\,>\,{\mu}/{2} \hskip 10mm  \Longrightarrow\hskip 10mm \mu\,<\, \nu+1\,.
\end{equation}

In the opposite limit, i.e. as the argument of the Legendre function in (\ref{u2}) approaches $-1$ from above ($z\to-\infty$), we can use connection formulae for the Legendre functions \citep{NIST:DLMF} to write the limit in terms of the limit $x\to1^{-}$ ($z\to\infty$). We can derive the formula
\begin{equation}
\lim_{x\to-1^{+}}{\mathrm P}_{\nu}^{\mu}(x)=\lim_{x\to1^{-}}\left[C_{\nu\mu}{\mathrm P}_{\nu}^{\mu}(x)-\frac{2}{\pi}S_{\nu\mu}{\mathrm Q}_{\nu}^{\mu}(x)\right],\label{connection}
\end{equation}
where $C_{\nu\mu}=\cos\left[(\nu+\mu)\pi\right]$, $S_{\nu\mu}=\sin\left[(\nu+\mu)\pi\right]$ and ${\mathrm Q}_{\nu}^{\mu}$ is an associated Legendre function of the second kind. Using (\ref{connection}), we have
\be
\label{lim_opposite}
\lim_{z\to-\infty}U_1(\bar{\varpi},z)=\lim_{z\to\infty}\left[\dfrac{C_{\nu\mu}{\mathrm P}_{\nu}^{\mu}\!\left(\dfrac{z}{\sqrt{\bar{\varpi}^2+z^2}}\right)-\dfrac{2}{\pi}S_{\nu\mu}{\mathrm Q}_{\nu}^{\mu}\!\left(\dfrac{z}{\sqrt{\bar{\varpi}^2+z^2}}\right)}{\biggl.\left(\bar{\varpi}^2+z^2\right)^{(\nu+1)/2}\biggr.}\right],\qquad
\ee

We have already determined the first term on the right-hand side of (\ref{lim_opposite}) (apart from a constant factor), i.e. the limit of $U_1$ as $z\to\infty$, which vanishes if $\mu<\nu+1$. We must now ensure that the second term, i.e.
\begin{equation}
-\frac{2}{\pi}S_{\nu\mu}\lim_{z\to\infty}\Bigg[\left(\bar{\varpi}^2+z^2\right)^{-(\nu+1)/2}{\mathrm Q}_{\nu}^{\mu}\left(\frac{z}{\sqrt{\bar{\varpi}^2+z^2}}\right)\Bigg].\label{Qlim}
\end{equation}
vanishes. In the limit $x\to1^{-}$, the function ${\mathrm Q}_{\nu}^{\mu}(x)$ has the asymptotic form
\begin{equation}
{\mathrm Q}_{\nu}^{\mu}(x)\sim\frac{\Gamma(-\mu)\Gamma(\nu+\mu+1)}{2\Gamma(\nu-\mu+1)}\left(\frac{1-x}{2}\right)^{\mu/2},
\end{equation}
which is valid if $\mu$ is not zero or a positive integer and if $\nu\pm\mu$ is not equal to a negative integer. By applying this formula to (\ref{Qlim}), we see that the limit is zero if $\mu>0$, but can be singular if $\mu<0$. In such a case, it is then clear that determining the limit in (\ref{Qlim}) requires us to determine the limit of $U_1$ for $z\to\infty$ (as described above), but with $\mu$ replaced by $-\mu$. This means that, to avoid any singular limits, we must have $-\mu<\nu+1$, i.e. $\mu>-(\nu+1)$.
 
For the second Legendre function in (\ref{laplace_sol_cyl}), we can follow a similar argument to the above and, by symmetry, arrive at the same restrictions on $\mu$ such that that part of the solution tends to zero in the limit $z\to\pm\infty$.

We have shown, therefore, that, given a particular value of $\nu$, the solution will tend to zero for $z\to\pm\infty$ provided we choose a value of $\mu$ in the range $-(\nu+1)<\mu<\nu+1$ that is not zero or a positive integer, and provided $\nu\pm\mu$ is not a negative integer. Since $U$ vanishes for $z\to\pm\infty$ under these conditions, the other terms in the force-balance equation (\ref{force-ampere-divB}a)
will also vanish, since they can be calculated from derivatives of $U$ (and so involve smaller powers of $z$ than $U$ does).

In this appendix, we have given an an example to illustrate the restrictions that arise when choosing $\mu$ and $\nu$ such that the solution is physically reasonable. It should not be considered as a general proof, however, since the asymptotic formulae we have used only apply to particular choices for $\mu$ and $\mu\pm\nu$.}\\




\begin{thebibliography}{34}
\providecommand{\natexlab}[1]{#1}
\markboth{F. Wilson and T. Neukirch}{3D solutions of the magnetohydrostatic equations}

\bibitem[\protect\citeauthoryear{{Abramowitz} and
  {Stegun}}{1964}]{Abramowitz-1964}
{Abramowitz}, M. and {Stegun}, I.A., {\itshape {Handbook of Mathematical
  Functions}},  1964 (National Bureau of Standards, Washington DC).

\bibitem[\protect\citeauthoryear{{Al-Salti} and
  {Neukirch}}{2010}]{Al-Salti-2010b}
{Al-Salti}, N. and {Neukirch}, T., {Three-dimensional solutions of the
  magnetohydrostatic equations: Rigidly rotating magnetized coronae in
  spherical geometry}. {\itshape Astron. Astrophys.}, 2010,
  \textbf{520}, A75.

\bibitem[\protect\citeauthoryear{{Al-Salti}
  {\itshape{et~al.}}}{2010}]{Al-Salti-2010a}
{Al-Salti}, N., {Neukirch}, T. and {Ryan}, R., {Three-dimensional solutions of
  the magnetohydrostatic equations: rigidly rotating magnetized coronae in
  cylindrical geometry}. {\itshape Astron. Astrophys.}, 2010,
  \textbf{514}, A38.

\bibitem[\protect\citeauthoryear{{Aulanier}
  {\itshape{et~al.}}}{1999}]{Aulanier-1999}
{Aulanier}, G., {D{\'e}moulin}, P., {Mein}, N., {van Driel-Gesztelyi}, L.,
  {Mein}, P. and {Schmieder}, B., {3-D magnetic configurations supporting
  prominences. III. Evolution of fine structures observed in a filament
  channel}. {\itshape Astron. Astrophys.}, 1999, \textbf{342},
  867--880.

\bibitem[\protect\citeauthoryear{{Bogdan} and {Low}}{1985}]{Bogdan-1985}
{Bogdan}, T.J. and {Low}, B.C., {Three-dimensional magnetostatic models of
  coronal structures.}; in {\itshape Bull. Am. Astron. Soc.}, Vol. ~17, Mar., 1985, p. 632.

\bibitem[\protect\citeauthoryear{Debnath}{2003}]{Debnath-2003}
Debnath, L., Recent applications of fractional calculus to science and
  engineering.. {\itshape Int. J. Math. Math. Sci.}, 2003, \textbf{2003}, 3413--3442.

\bibitem[\protect\citeauthoryear{{Engheta}}{1996}]{Engheta-1996}
{Engheta}, N., {On fractional calculus and fractional multipoles in
  electromagnetism}. {\itshape IEEE T. Antenn. Propag.},
  1996, \textbf{44}, 554--566.

\bibitem[\protect\citeauthoryear{{Gibson} and {Bagenal}}{1995}]{Gibson-1995}
{Gibson}, S.E. and {Bagenal}, F., {Large-scale magnetic field and density
  distribution in the solar minimum corona}. {\itshape J. Geophys. Res.}, 1995, \textbf{100}, 19865--19880.

\bibitem[\protect\citeauthoryear{{Gibson}
  {\itshape{et~al.}}}{1996}]{Gibson-1996}
{Gibson}, S.E., {Bagenal}, F. and {Low}, B.C., {Current sheets in the solar
  minimum corona}. {\itshape J. Geophys. Res.}, 1996,
  \textbf{101}, 4813--4824.

\bibitem[\protect\citeauthoryear{{Lanza}}{2008}]{Lanza-2008}
{Lanza}, A.F., {Hot Jupiters and stellar magnetic activity}. {\itshape
  Astron. Astrophys.}, 2008, \textbf{487}, 1163--1170.

\bibitem[\protect\citeauthoryear{{Lanza}}{2009}]{Lanza-2009}
{Lanza}, A.F., {Stellar coronal magnetic fields and star-planet interaction}.
  {\itshape Astron. Astrophys.}, 2009, \textbf{505}, 339--350.

\bibitem[\protect\citeauthoryear{{Low}}{1982}]{low-1982}
{Low}, B.C., {Magnetostatic atmospheres with variations in three dimensions}.
  {\itshape Astrophys. J.}, 1982, \textbf{263}, 952--969.

\bibitem[\protect\citeauthoryear{{Low}}{1984}]{low-1984}
{Low}, B.C., {Three-dimensional magnetostatic atmospheres - Magnetic field with
  vertically oriented tension force}. {\itshape Astrophys. J.},
  1984, \textbf{277}, 415--421.

\bibitem[\protect\citeauthoryear{{Low}}{1985}]{low-1985}
{Low}, B.C., {Three-dimensional structures of magnetostatic atmospheres. I -
  Theory}. {\itshape Astrophys. J.}, 1985, \textbf{293}, 31--43.

\bibitem[\protect\citeauthoryear{{Low}}{1991}]{low-1991}
{Low}, B.C., {Three-dimensional structures of magnetostatic atmospheres. III -
  A general formulation}. {\itshape Astrophys. J.}, 1991,
  \textbf{370}, 427--434.

\bibitem[\protect\citeauthoryear{{Low}}{1992}]{low-1992}
{Low}, B.C., {Three-dimensional structures of magnetostatic atmospheres. IV -
  Magnetic structures over a solar active region}. {\itshape Astrophys. J.}, 1992, \textbf{399}, 300--312.

\bibitem[\protect\citeauthoryear{{Low}}{1993{\natexlab{a}}}]{low-1993a}
{Low}, B.C., {Three-dimensional structures of magnetostatic atmospheres. V -
  Coupled electric current systems}. {\itshape Astrophys. J.},
  1993{\natexlab{a}}, \textbf{408}, 689--692.

\bibitem[\protect\citeauthoryear{{Low}}{1993{\natexlab{b}}}]{low-1993b}
{Low}, B.C., {Three-dimensional structures of magnetostatic atmospheres. VI -
  Examples of coupled electric current systems}. {\itshape Astrophys. J.}, 1993{\natexlab{b}}, \textbf{408}, 693--706.

\bibitem[\protect\citeauthoryear{{Low}}{2005}]{low-2005}
{Low}, B.C., {Three-dimensional Structures of Magnetostatic Atmospheres. VII.
  Magnetic Flux Surfaces and Boundary Conditions}. {\itshape Astrophys. J.}, 2005, \textbf{625}, 451--462.

\bibitem[\protect\citeauthoryear{{MacTaggart}
  {\itshape{et~al.}}}{2016}]{Mactaggart-2016}
{MacTaggart}, D., {Gregory}, S.G., {Neukirch}, T. and {Donati}, J.F.,
  {Magnetohydrostatic modelling of stellar coronae}. {\itshape Mon. Not. R. Astron. Soc.}, 2016, \textbf{456}, 767--774.

\bibitem[\protect\citeauthoryear{Mestel}{2003}]{mestel2003}
Mestel, L., {\itshape Stellar Magnetism}, International Series of Monographs on
  Physics 2003 (Clarendon Press).

\bibitem[\protect\citeauthoryear{{Neukirch}}{1995}]{Neukirch-1995}
{Neukirch}, T., {On self-consistent three-dimensional analytic solutions of the
  magnetohydrostatic equations.}. {\itshape Astron. Astrophys.}, 1995,
  \textbf{301}, 628.

\bibitem[\protect\citeauthoryear{{Neukirch}}{1997}]{Neukirch-1997}
{Neukirch}, T., {Nonlinear self-consistent three-dimensional arcade-like
  solutions of the magnetohydrostatic equations.}. {\itshape Astron. Astrophys.}, 1997, \textbf{325}, 847--856.

\bibitem[\protect\citeauthoryear{{Neukirch}}{2009}]{Neukirch-2009b}
{Neukirch}, T., {Three-dimensional analytical magnetohydrostatic equilibria of
  rigidly rotating magnetospheres in cylindrical geometry}. {\itshape
  Geophys. Astrophys. Fluid Dyn.}, 2009, \textbf{103}, 535--547.

\bibitem[\protect\citeauthoryear{{Neukirch} and
  {Rast{\"a}tter}}{1999}]{Neukirch-1999}
{Neukirch}, T. and {Rast{\"a}tter}, L., {A new method for calculating a special
  class of self-consistent three-dimensional magnetohydrostatic equilibria}.
  {\itshape Astron. Astrophys.}, 1999, \textbf{348}, 1000--1004.

\bibitem[\protect\citeauthoryear{Olver {\itshape{et~al.}}}{2010}]{NIST:DLMF}
Olver, F.W., Lozier, D.W., Boisvert, R.F. and Clark, C.W., {\itshape NIST
  Handbook of Mathematical Functions}, 1st edn,  2010 (New York, NY, USA:
  Cambridge University Press).

\bibitem[\protect\citeauthoryear{{Petrie} and {Neukirch}}{2000}]{Petrie-2000}
{Petrie}, G.J.D. and {Neukirch}, T., {The Green's function method for a special
  class of linear three-dimensional magnetohydrostatic equilibria}. {\itshape
  Astron. Astrophys.}, 2000, \textbf{356}, 735--746.

\bibitem[\protect\citeauthoryear{{Ruan} {\itshape{et~al.}}}{2008}]{Ruan-2008}
{Ruan}, P., {Wiegelmann}, T., {Inhester}, B., {Neukirch}, T., {Solanki}, S.K.
  and {Feng}, L., {A first step in reconstructing the solar corona
  self-consistently with a magnetohydrostatic model during solar activity
  minimum}. {\itshape Astron. Astrophys.}, 2008, \textbf{481},
  827--834.

\bibitem[\protect\citeauthoryear{{Rudenko}}{2001}]{Rudenko-2001}
{Rudenko}, G.V., {A constructing method for a self-consistent three-dimensional
  solution of the magnetohydrostatic equations using full-disk magnetogram
  data}. {\itshape Sol. Phys.}, 2001, \textbf{198}, 279--287.

\bibitem[\protect\citeauthoryear{{Wiegelmann}
  {\itshape{et~al.}}}{2017}]{Wiegelmann-2017}
{Wiegelmann}, T., {Neukirch}, T., {Nickeler}, D.H., {Solanki}, S.K., {Barthol},
  P., {Gandorfer}, A., {Gizon}, L., {Hirzberger}, J., {Riethm{\"u}ller}, T.L.,
  {van Noort}, M., {Blanco Rodr{\'{\i}}guez}, J., {Del Toro Iniesta}, J.C.,
  {Orozco Su{\'a}rez}, D., {Schmidt}, W., {Mart{\'{\i}}nez Pillet}, V. and
  {Kn{\"o}lker}, M., {Magneto-static Modeling from Sunrise/IMaX: Application to
  an Active Region Observed with Sunrise II}. {\itshape Astrophys. J. Supplement Series}, 2017, \textbf{229}, 18.

\bibitem[\protect\citeauthoryear{{Wiegelmann}
  {\itshape{et~al.}}}{2015}]{Wiegelmann-2015}
{Wiegelmann}, T., {Neukirch}, T., {Nickeler}, D.H., {Solanki}, S.K.,
  {Mart{\'{\i}}nez Pillet}, V. and {Borrero}, J.M., {Magneto-static Modeling of
  the Mixed Plasma Beta Solar Atmosphere Based on Sunrise/IMaX Data}. {\itshape
  Astrophys. J.}, 2015, \textbf{815}, 10.

\bibitem[\protect\citeauthoryear{{Zhao} and {Hoeksema}}{1993}]{Zhao-1993}
{Zhao}, X. and {Hoeksema}, J.T., {Unique determination of model coronal
  magnetic fields using photospheric observations}. {\itshape Sol. Phys.},
  1993, \textbf{143}, 41--48.

\bibitem[\protect\citeauthoryear{{Zhao} and {Hoeksema}}{1994}]{Zhao-1994}
{Zhao}, X. and {Hoeksema}, J.T., {A coronal magnetic field model with
  horizontal volume and sheet currents}. {\itshape Sol. Phys.}, 1994,
  \textbf{151}, 91--105.

\bibitem[\protect\citeauthoryear{{Zhao} {\itshape{et~al.}}}{2000}]{Zhao-2000}
{Zhao}, X.P., {Hoeksema}, J.T. and {Scherrer}, P.H., {Modeling the 1994 April
  14 Polar Crown SXR Arcade Using Three-Dimensional Magnetohydrostatic
  Equilibrium Solutions}. {\itshape Astrophys. J.}, 2000,
  \textbf{538}, 932--939.

\end{thebibliography}
\end{document}